\documentclass[sigconf, pbalance]{acmart}

\usepackage{iftex}
\ifPDFTeX
  \usepackage[utf8]{inputenc}
  \usepackage[T1]{fontenc}
\fi
\usepackage{graphicx}
\usepackage{subcaption}
\usepackage{booktabs}
\usepackage{multirow}
\usepackage{array}
\usepackage{xcolor}
\usepackage{colortbl}
\usepackage{enumitem}
\usepackage{xspace}
\usepackage{microtype}
\usepackage{url}
\usepackage{multicol}
\usepackage{tablefootnote}
\usepackage{adjustbox}
\usepackage{needspace}
\usepackage{pict2e}
\usepackage{algorithm}
\usepackage{algpseudocode}
\algrenewcommand\algorithmicrequire{\textbf{Input:}}

\definecolor{decobg}{HTML}{F9EFEA}
\definecolor{basegray}{HTML}{707070}
\newcommand{\pmsd}[1]{\,{\scriptsize$\pm$#1}}
\newcommand{\xs}{{\scriptsize$\times$}}
\DeclareRobustCommand{\mkboth}{\raisebox{.3ex}{\setlength{\unitlength}{1pt}\begin{picture}(3,3)\polygon*(0,0)(3,0)(1.5,3)\end{picture}}}
\DeclareRobustCommand{\mkteach}{\raisebox{.3ex}{\setlength{\unitlength}{1pt}\begin{picture}(3,3)\linethickness{.3pt}\polygon(0,0)(3,0)(1.5,3)\end{picture}}}

\setlist{nosep,leftmargin=*}

\newcommand{\deco}{Deco\xspace}
\newcommand{\decoflat}{Deco-flat\xspace}
\newcommand{\fusedgr}{FusedGR\xspace}
\newcommand{\flatgr}{FlatGR\xspace}
\newcommand{\flatgrmatch}{FlatGR-match\xspace}
\newcommand{\decoce}{Deco-CE\xspace}
\newcommand{\refslash}{\setbox0\hbox{\xs0.0\,/\,\xs0.0}\makebox[\wd0][c]{\footnotesize ref.}}

\begin{document}
\title[Separating Thinking from Spelling]{Separating Thinking from Spelling for Efficient \\ Generative Recommendation}
\author{Danil Gusak, Anna Volodkevich, Ivan Oseledets, Evgeny Frolov}

\copyrightyear{2026}
\acmYear{2026}
\setcopyright{cc}
\setcctype{by}



\begin{teaserfigure}
  \setlength{\abovecaptionskip}{6pt}
  \centering
  \includegraphics[width=\textwidth]{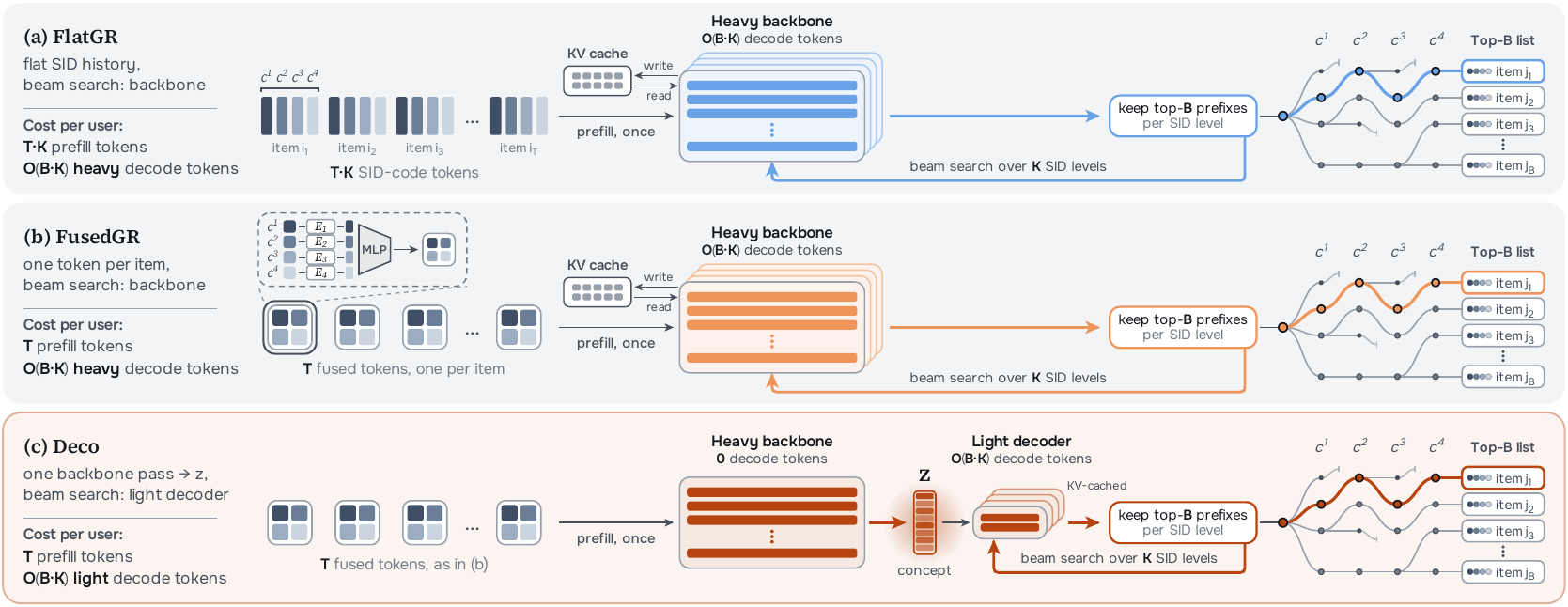}
  \caption{Compressing the input sequence and delegating beam search to a light decoder. For a history of \(T\) items with \(K\) SID codes each, \flatgr (a) reads \(TK\) history tokens, while \fusedgr (b) and \deco (c) read \(T\) fused tokens. With beam width \(B\), \flatgr and \fusedgr process \(O(BK)\) decode tokens through the heavy backbone. \deco computes a compact concept \(z\) in one backbone pass and delegates beam search to a light decoder conditioned on \(z\) and the generated prefix.}
  \label{fig:teaser}
\end{teaserfigure}

\begin{abstract}
Generative recommenders retrieve items by writing their semantic IDs (SIDs) code by code under trie-constrained beam search, an approach now deployed on large web platforms.
Standard autoregressive interfaces pay for the SID length $K$ twice: a history of $T$ items enters the backbone as $TK$ code tokens, and every beam-search step runs through the backbone, so returning $B$ candidates costs $O(BK)$ backbone decode tokens per user.
The second cost often dominates, and it grows with the backbone, the SID depth and the candidate budget.

We introduce \deco (decoding from concepts), which separates reading the history from writing the identifier.
The backbone reads each history item as one fused token and condenses the history once into a \emph{compact next-item concept}; a light KV-cached decoder conditioned on the concept and the generated prefix runs the entire beam search, while the identifiers, the trie and the autoregressive factorization stay unchanged.
The decoder approaches full-backbone accuracy with cross-entropy alone, and distillation improves it further.
On seven datasets from e-commerce, movies and music streaming, \deco achieves comparable or better \mbox{accuracy} than both a standard generative recommender and the \mbox{full-backbone} teacher with fused inputs, while delivering 3.2--8.6$\times$ their throughput under a 50\,ms p95 latency budget in a production-style serving pipeline.
Delegating beam search makes larger candidate pools and models practical: \deco generates 1000 candidates faster than its teacher generates 100, and with a 1.5B-parameter LLM backbone, it matches the 69M-parameter teacher's throughput while achieving higher accuracy.

\end{abstract}

\maketitle

\section{Introduction}
\label{sec:intro}
Recommendation shapes how people find products, films and music on the Web, and every request begins with candidate retrieval: selecting a few hundred to a few thousand items from a large catalog within tens of milliseconds.
Generative retrieval casts this stage as sequence generation~\cite{tay2022transformer,rajput2023tiger}.
Each item receives a semantic ID (SID), a short tuple of discrete codes derived from its content; SIDs let similar items share code prefixes and keep the output vocabulary far smaller than the catalog.
A sequence model reads the user's history and generates SIDs code by code using beam search constrained by a trie of valid SIDs.
The SIDs retained at the final step identify the candidate items.
Generative recommenders (GR) inherit the architectures and scaling practices of language models~\cite{zheng2024lcrec,liu2025understanding,zhai2024actions}, already serve industrial traffic~\cite{deng2025onerec,zhou2025onerecv2} and increasingly build on pretrained LLMs with billions of parameters~\cite{he2025plum,kong2025minionerec}.

SID-based retrieval has two sources of inference cost: encoding the history and decoding candidates.
In a flat representation (\flatgr, Figure~\ref{fig:teaser}a), a history of $T$ items with $K$ codes each occupies $TK$ tokens.
Fusing each item's code embeddings into one token (\fusedgr, Figure~\ref{fig:teaser}b) reduces the input to $T$ tokens.
Both models encode the history once during \emph{prefill}, but use the full backbone for autoregressive (AR) decoding.
Beam search of width $B$ over $K$-level SIDs requires $O(BK)$ backbone decode tokens per user, each attending to the cached history.
In our measurements at beam width 100, beam search takes up to 91\% of generation time, even with a beam-shared history cache (\S\ref{sec:rq2}).
This cost grows with the SID depth, the candidate budget and the backbone size, so input compression alone does not remove this decoding bottleneck.

Our starting point is to separate modeling the user's preferences from spelling out candidate SIDs.
\flatgr and \fusedgr use the same backbone to represent the history and score the next code for every candidate prefix, spending $O(BK)$ full-backbone decode tokens per user.
Our experiments show that a compact next-item representation from the backbone lets a much smaller decoder generate SIDs \emph{without sacrificing recommendation quality} (\S\ref{sec:rq1}).
Preference modeling can thus remain in the backbone, while a light decoder handles every step of candidate generation.

We propose \deco (\textbf{De}coding from \textbf{co}ncepts), which makes this division explicit (Figure~\ref{fig:teaser}c).
The backbone reads the fused history once and emits a compact \emph{next-item concept} shared by every candidate in the beam.
It delegates \emph{the entire trie-constrained beam search} to a light KV-cached decoder conditioned only on this concept and the generated prefix.
The SID codes and trie stay unchanged.
The decoder can be trained with cross-entropy alone, and distillation from a full-backbone model improves its accuracy.
For repeat-heavy domains such as music streaming, a light pointer over the history retrieves the exact items a user returns to.
The light decoder returns final candidates without the backbone verification used in speculative decoding~\cite{lin2025efficient}, and unlike parallel or single-pass decoders~\cite{hou2025rpg,lin2025order}, it keeps every step autoregressive and trie-constrained.

Our contributions are as follows:
\begin{itemize}[leftmargin=*,nosep]
    \item We show that SID-based generative retrieval often spends more time spelling identifiers than reading the history: beam-search decoding dominates prefill in both \flatgr and \fusedgr. Fusing history items reduces prefill cost, but beam search still requires $O(BK)$ full-backbone decode tokens per user.
    \item We propose \deco, which combines fused history representations with delegated decoding. The backbone produces a compact next-item concept, and a light decoder runs the entire trie-constrained beam search using only this concept and the generated prefix. Training supervises every next-item transition in a single \mbox{backbone} pass.
    \item On seven datasets, concept decoding matches or exceeds the accuracy of the corresponding full-backbone baselines for both flat and fused histories. In a production-style pipeline, \deco delivers 2.8--5.1$\times$ \fusedgr's throughput at batch size 128 and 3.2--7.3$\times$ under a 50\,ms p95 latency budget.
    \item We show that the speedup grows with the candidate budget, SID depth and backbone size. \deco extends to pretrained LLM backbones, reaching a 16$\times$ speedup over \fusedgr with a 3B model.
\end{itemize}

\section{Related Work}
\label{sec:related}
\paragraph{Generative recommendation with semantic IDs.}
Generative retrieval represents each document or item by a short sequence of discrete tokens and retrieves it by decoding this sequence~\cite{tay2022transformer,sun2023learning}.
TIGER~\cite{rajput2023tiger} brought this idea to sequential recommendation with SIDs from residual quantization~\cite{zeghidour2022soundstream,lee2022autoregressive}.
Later work improved item tokenization~\cite{wang2024letter,zhu2024cost,hou2025actionpiece,liu2025etegrec,fu2026differentiable,liang2026rethinking}, combined generative and dense retrieval~\cite{yang2025liger}, adapted pretrained LLMs to SID-based recommendation~\cite{zheng2024lcrec,he2025plum,kong2025minionerec,denadai2026neo}, and scaled generative recommenders to industrial traffic~\cite{deng2025onerec,zhou2025onerecv2}.
These works develop item representations and recommendation models; we focus on the computation repeated during candidate decoding.
Full-backbone SID recommenders, whether they read the history as a flat sequence of SID code tokens or as one fused token per item~\cite{10.1145/3805712.3808437}, generate the codes either in the backbone or, as in encoder--decoder models such as TIGER and OneRec~\cite{deng2025onerec}, in a decoder of comparable or larger size that cross-attends to the encoded history at every step.
OneRec-V2~\cite{zhou2025onerecv2} drops the encoder and lets a large decoder read the context through shared keys and values, cutting the cost of reading; writing, which we target, still runs through the large model.
As generative recommenders grow, adopt large decoder-only LLMs, and reach industrial scale, reducing repeated decoding work becomes important for meeting latency budgets.

\paragraph{Efficient decoding for generative recommendation.}
The cost of beam search is a recognized obstacle to deployment~\cite{lin2025efficient,volodkevich2025autoregressive}.
Serving optimizations include early sorting termination, valid-item filtering and overlap between computation and data transfers~\cite{sun2025xgr}.
Speculative decoding verifies a small model's drafts with the backbone~\cite{lin2025efficient}; NEZHA uses an autoregressive head over per-level backbone states and filters invalid SIDs afterwards~\cite{wang2026nezha}.
RPG generates unordered codes in parallel and searches a graph~\cite{hou2025rpg}, \mbox{SETRec} generates order-agnostic identifiers~\cite{lin2025order}, and LLaDA-Rec uses discrete diffusion~\cite{shi2025lladarec}.
\deco instead separates reading from writing: it runs the entire autoregressive beam search in a light decoder conditioned on a fixed-size, cacheable next-item concept produced by the heavy backbone.
Wider beams and deeper SIDs increase decoder computation without further backbone calls.
Unlike the parallel and diffusion-based methods above, \deco retains left-to-right autoregressive decoding with trie constraints, ensuring that every retained prefix extends to a catalog item.
Among the generative baselines evaluated in Table~\ref{tab:baselines}, \deco achieves the highest recommendation accuracy on both datasets and higher throughput than the autoregressive models.

\paragraph{Global--local generation, distillation and copying.}
Global--local language models use a global model over patches or blocks and a local token decoder~\cite{yu2023megabyte,ho2024block}; related designs appear in deep-encoder, shallow-decoder translation~\cite{kasai2021deep} and concept-level language modeling~\cite{barrault2024lcm,dragunov2025sonar}.
\deco applies this pattern to SID retrieval, where returning many candidates typically requires wider beams than in LLM text generation~\cite{sun2025xgr}.
Sharing one concept across the candidate prefixes keeps this repeated work in the light decoder.
To improve retrieval accuracy, we use distillation~\cite{hinton2015distilling,kim2016sequence,tang2018ranking}, which can let equal-capacity students surpass their teachers~\cite{furlanello2018born} and acts as teacher-informed label smoothing in our level-wise objective~\cite{yuan2020revisiting}.
To handle repeat consumption, we add a history pointer, following copy mechanisms~\cite{vinyals2015pointer,see2017get} and repeat-aware recommendation~\cite{ren2019repeatnet}; its mixture with the generator factorizes over the SID trie.

\section{Method}
\label{sec:method}
\subsection{Semantic-ID Generation and Its Cost}
\label{sec:prelim}

\paragraph{Semantic IDs.}
A tokenizer maps every item $i$ to a semantic ID $s(i)=(c_i^1,\dots,c_i^K)$ of $K$ codes from $\{1,\dots,V\}$ by residual quantization of a content embedding, so the first code places the item in a coarse cluster and later codes refine it~\cite{rajput2023tiger,zhou2025openonerec}.
Catalog SIDs form a prefix tree $\mathcal T$, and $\mathcal T(\pi)$ is the set of valid next codes after a prefix $\pi$.
Given a history $H=(i_1,\dots,i_T)$, a generative recommender factorizes the next-item distribution over code levels,
\begin{equation}
p\big(s(i)\mid H\big)=\prod_{k=1}^{K} p\big(c_i^k \mid c_i^{<k}, H\big),
\label{eq:factorization}
\end{equation}
and retrieves candidates by beam search of width $B$ over $\mathcal T$: at every level, each surviving prefix is extended by its valid codes, and the $B$ best extensions are kept.

\paragraph{Flat autoregressive decoding.}
\flatgr represents the history as a flat sequence of $TK$ code tokens, following TIGER~\cite{rajput2023tiger} and many subsequent SID models. We use a decoder-only backbone~\cite{zhou2025onerecv2,he2025plum} to read this sequence and autoregressively generate the next item's $K$-code SID.
The prefill encodes the $L_{\mathrm{in}}$ history tokens once and caches their keys and values.
Beam search then reads the first level from the prefill output, and each of the other $K-1$ levels passes $B$ new tokens through the backbone, each attending to the cached history.
To leading order, the compute per user, which bounds throughput under load, is
\begin{equation}
C_{\mathrm{AR}} \approx F_{\mathrm{bb}}(L_{\mathrm{in}}) + B\,(K-1)\,D_{\mathrm{bb}}(L_{\mathrm{in}}),
\label{eq:cost-ar}
\end{equation}
where $F_{\mathrm{bb}}$ is the cost of the prefill and $D_{\mathrm{bb}}$ is the cost of one backbone decode token, which grows with the history length it attends to.
For \flatgr, $L_{\mathrm{in}}=TK$: the model pays for the SID length twice, in what it reads and, multiplied by the beam, in what it writes.

\subsection{Fused History}
\label{sec:fused}

A flat history represents each past item with $K$ code tokens.
Following prior work on item-level SID representations~\cite{hou2025rpg,10.1145/3805712.3808437}, \fusedgr instead uses one token per history item.
A small MLP, the \emph{fuser}, combines the code embeddings into a single item vector:
\begin{equation}
x_t=\mathrm{MLP}\big(\big[E_1(c^1_{i_t});\,\dots;\,E_K(c^K_{i_t})\big]\big).
\label{eq:fuser}
\end{equation}
This shortens the input from $TK$ to $T$ tokens.
Dense training is native to a flat model, since every code token is a next-token target; a fused input contains no code tokens, so the codes of the next item must be supplied.
On a training sequence $(i_1,\dots,i_{T+1})$, we append, for every position $t$, the gold prefix of $i_{t+1}$ under a block-causal mask, so that its tokens attend only to $x_1,\dots,x_t$ and causally to one another.
One forward pass over $TK$ tokens, as in a flat model, shares the history computation across targets and evaluates
\begin{equation}
\mathcal L_{\mathrm{AR}}=-\sum_{t=1}^{T}\sum_{k=1}^{K}\log p_\theta\big(c^k_{i_{t+1}}\mid c^{<k}_{i_{t+1}},\,x_1,\dots,x_t\big).
\label{eq:dense}
\end{equation}
Fusion, however, changes only what the backbone reads: with $L_{\mathrm{in}}=T$ in Eq.~\eqref{eq:cost-ar}, the $O(BK)$ decode tokens remain.

\subsection{Decoding from Concepts}
\label{sec:deco}

We separate user modeling from SID generation: the backbone produces a next-item concept from the history, and a light decoder generates SIDs conditioned on this concept.
With the flat history of \flatgr, \decoflat isolates the effect of delegating beam search.
With fused history, \deco also reduces prefill cost (Figure~\ref{fig:teaser}c).
The backbone $f_\theta$ of width $d$ reads the history once, and a linear map turns its last state $h_T$ (for a flat history, the state at the last code token) into a \emph{next-item concept} of one or a few vectors ($M{=}2$ in our experiments; $M$ from 1 to 8 performs similarly):
\begin{equation}
z = \operatorname{reshape}\big(W_z\, h_T\big) \in \mathbb R^{M\times d}.
\label{eq:concept}
\end{equation}
The concept does not commit to one item: it conditions a distribution over plausible next items, which a small causal transformer $g_\phi$ enumerates code by code.
The decoder reads the concept, mapped to its width by an MLP adapter $u=\psi(z)$, followed by its own level-specific embeddings $e_j(c^j)$ of the codes generated so far.
Level-wise heads $W_k$ then score the valid children of the prefix:
\begin{equation}
\begin{aligned}
p_\phi\big(\cdot\mid c^{<k},z\big) &= \operatorname{softmax}_{\mathcal T(c^{<k})}\big(W_k\, o_k\big),\\
o_k &= g_\phi\big(u_1,\dots,u_M,\,e_1(c^1),\dots,e_{k-1}(c^{k-1})\big)_{\mathrm{last}}.
\end{aligned}
\label{eq:decoder}
\end{equation}
The concept is the only interface between the two stages: the decoder \emph{never attends to the history} and keeps no backbone cache.
Beam search therefore runs entirely inside $g_\phi$ with its own KV cache (Algorithm~\ref{alg:deco}).
Ignoring trie masking and top-$B$ selection, which all models share, the compute per user is
\begin{equation}
C_{\mathrm{Deco}} \approx F_{\mathrm{bb}}(T) + F_{\mathrm{dec}}(M) + B\,(K-1)\,D_{\mathrm{dec}}(M{+}K{-}1),
\label{eq:cost-deco}
\end{equation}
where $F_{\mathrm{dec}}$ and $D_{\mathrm{dec}}$ are the prefill and decode costs of the decoder (by default two layers of width 256, against the backbone's eight of width 640), which attends to at most $M+K-1$ tokens.
A larger backbone or a longer history increases \deco's one-time encoding cost, while wider beams and deeper SIDs increase computation in the light decoder.
As the beam width $B$ grows for a fixed $K$, the compute speedup approaches
\begin{equation}
\frac{C_{\mathrm{AR}}}{C_{\mathrm{Deco}}}\;\longrightarrow\;\frac{D_{\mathrm{bb}}(L_{\mathrm{in}})}{D_{\mathrm{dec}}(M{+}K{-}1)},
\label{eq:speedup}
\end{equation}
a ratio that grows with the backbone size and the history length; a longer history, however, mainly lengthens the prefill, which both pay once.

\begin{algorithm}[t]
\caption{Retrieval with \deco for one user}
\label{alg:deco}
\begin{algorithmic}[1]
\small
\Require history $(i_1,\dots,i_T)$, beam width $B$, SID trie $\mathcal T$
\State $x_t \gets \mathrm{MLP}\big([E_1(c^1_{i_t});\dots;E_K(c^K_{i_t})]\big)$ for $t=1,\dots,T$ \Comment{Eq.~\eqref{eq:fuser}}
\State $z \gets \operatorname{reshape}(W_z f_\theta(x_{1:T})_T)$,\quad $u\gets \psi(z)$ \Comment{the only backbone pass}
\State prefill $g_\phi$ with $u_1,\dots,u_M$;\quad $\mathcal B\gets\{(\varnothing,\,0)\}$
\For{$k=1,\dots,K$}
  \State $\mathcal B\gets \operatorname{top}_B\big\{(\pi c,\ r+\log p_\phi(c\mid\pi,z)) : (\pi,r)\in\mathcal B,\ c\in\mathcal T(\pi)\big\}$
  \If{$k<K$}
    \State feed the last code of each beam to $g_\phi$, reusing the decoder KV cache of its parent
  \EndIf
\EndFor
\State \Return the SIDs in $\mathcal B$ ordered by score $r$, resolved to items
\end{algorithmic}
\end{algorithm}

\paragraph{The concept as a serving state.}
Since the decoder sees the user only through $z$, its adapted form $u=\psi(z)$ is a complete per-user state for the generator: 1~KiB in bf16 for any backbone, against 0.4--7.8~MiB for the history KV cache of a full-backbone model (Figure~\ref{fig:per-user-state}, Appendix~\ref{app:results}).
This state can be computed in advance when the history changes, leaving only the light decoder online, although our timings include recomputation for every request.

\subsection{Training the Light Decoder}
\label{sec:kd}

\paragraph{Learning from data.}
The backbone, the fuser and the decoder are trained jointly and densely: a concept $z_t$ is read at every position $t$ (Eq.~\eqref{eq:concept} at $h_t$), and the decoder predicts the codes $c_\star$ of $i_{t+1}$ for all $t$ in parallel, as $T$ short sequences of $M+K-1$ tokens.
With cross-entropy alone, $\mathcal L_{\mathrm{CE}}=-\sum_t\sum_k\log p^k_\phi(c^k_\star)$, where $p^k_\phi=p_\phi(\cdot\mid c_\star^{<k},z_t)$, \deco needs nothing beyond interaction data; we call this teacher-free variant \decoce.
A full-backbone model can provide a distribution over valid continuations, giving the light decoder richer supervision than hard next-item labels alone.

\paragraph{Learning from a full-backbone model.}
To further improve retrieval quality, we regularize the decoder with knowledge distillation (KD)~\cite{hinton2015distilling,yuan2020revisiting}.
Frozen full-backbone teachers attend to the history at each code level: \fusedgr for \deco and \flatgr for \decoflat.
The decoder learns from both the observed code and the teacher's next-code distribution $q^k$ along the gold prefix $c_\star^{<k}$:
\begin{equation}
\mathcal L=\sum_{t=1}^{T}\sum_{k=1}^{K}\Big[-w_{\mathrm{ce}}\log p^k_\phi\big(c^k_\star\big)+w_{\mathrm{kd}}\,\tau^2\,\mathrm{KL}\big(q^k_\tau\,\big\Vert\,p^k_{\phi,\tau}\big)\Big],
\label{eq:loss}
\end{equation}
where $\tau$ is a temperature, both distributions are restricted to $\mathcal T(c_\star^{<k})$, and the weights $w_{\mathrm{ce}}$ and $w_{\mathrm{kd}}$ are fixed once from the initial loss values (Appendix~\ref{app:impl}).

With cross-entropy $\ell(y,p)=-\sum_c y_c\log p_c$ and $\tau=1$, since $\mathrm{KL}(q\,\Vert\,p)=\ell(q,p)-\ell(q,q)$, the objective is, up to a constant,
\begin{equation}
\mathcal L=(w_{\mathrm{ce}}+w_{\mathrm{kd}})\sum_{t=1}^{T}\sum_{k=1}^{K}\ell\Big(\alpha\,q^k+(1-\alpha)\,\delta_{c^k_\star},\;p^k_\phi\Big),
\label{eq:interp}
\end{equation}
with $\alpha=w_{\mathrm{kd}}/(w_{\mathrm{ce}}+w_{\mathrm{kd}})$: a single cross-entropy against a target that mixes the teacher's belief with the observed code.
Distillation thus acts as teacher-informed label smoothing~\cite{yuan2020revisiting}: instead of spreading uniformly over all codes, the smoothing mass goes to the trie-valid continuations of the same gold prefix that the teacher finds plausible.
The observed code keeps a point mass of $1-\alpha$, so the student is guided by the teacher without being bound to it where the two disagree; the teacher is needed only for training.

\subsection{Copying from the History}
\label{sec:copy}

A concept summarizes what comes next, not which past items the user has seen.
In domains such as music streaming, however, users often return to items they have consumed before~\cite{ren2019repeatnet,repeatconsumption2020doi}.
Such a target is easy to point to but hard to generate: the model must know \emph{which} past item to repeat, and a fixed-size concept is not meant to carry the identities of a hundred earlier items.
Following pointer-generator networks~\cite{vinyals2015pointer,see2017get}, we add a copy pointer over the history as a second distribution over SIDs,
\begin{equation}
P(s\mid H)=(1-\lambda)\,p_\phi(s\mid z)+\lambda\sum_{t:\,s(i_t)=s} a_t,
\label{eq:copy}
\end{equation}
where $a_t$ is the softmax attention weight from the last backbone state $h_T$ to the state of history item $t$, and $\lambda=\operatorname{sigmoid}(w_\lambda^{\top}h_T)$ is a learned gate.
The pointer is a single attention read over the history, like one cross-attention head, except that its weights are used directly as a distribution over past items rather than to mix their values.
Both components live in the same SID space, so the mixture factorizes over the trie exactly.
With $p_\phi(\pi)=p_\phi(\pi\mid z)$ the generator probability of a prefix $\pi$ and $A(\pi)$ the attention mass of the history items whose SID starts with $\pi$,
\begin{equation}
P\big(c\mid \pi\big)=\frac{(1-\lambda)\,p_\phi(\pi c)+\lambda\,A(\pi c)}{(1-\lambda)\,p_\phi(\pi)+\lambda\,A(\pi)}.
\label{eq:copy-trie}
\end{equation}
Beam search thus runs unchanged on the mixture, and training maximizes its exact likelihood, with the distillation term of Eq.~\eqref{eq:loss} applied to the full mixture.
The pointer is computed once from the backbone states and keeps only the history SIDs, their attention weights and the gate, so the decoder still never attends to the history.
We use this pointer with both the full-backbone baselines and \deco on Yambda (\S\ref{sec:rq4}).

\section{Experiments}
\label{sec:experiments}
Our experiments address the following research questions:
\begin{enumerate}[font={\bfseries}, label={RQ\arabic*}, wide, labelindent=0pt, before=\vspace{3pt}]
    \item Does delegating the entire beam search to a light decoder preserve recommendation accuracy?
    \item Where does generation time go, and how does delegating beam search affect serving latency and throughput?
    \item How do the gains scale with the candidate budget, the SID depth, the history length and the backbone size?
    \item Is a compact concept a sufficient interface to the history, and where does it fall short?
    \item How large does the light decoder need to be, and how should it be trained?
\end{enumerate} \label{rqs}

\subsection{Experimental Setup}
\label{sec:setup}

\subsubsection{Datasets}
\label{sec:datasets}

We use seven public datasets that differ in domain, scale, history length and repeat behavior (Table~\ref{tab:dataset_stats}).
Toys, Beauty and Sports come from Amazon Reviews 2014~\cite{mcauley2015imagebased}, and BPC-23 (Beauty and Personal Care) and Books-23 from the much larger Amazon Reviews 2023~\cite{hou2024bridging}.
We use their published 5-core subsets as implicit feedback and describe items by their product metadata.
MovieLens is a subset of MovieLens-25M~\cite{harper2015movielens}, with movies described by title, genres, directors, actors, writers and plot.
Yambda~\cite{yambda} is a music-streaming log: we take the last 30 days of Yambda-500M listens, remove exact duplicates, collapse consecutive repeats of a track~\cite{10.1145/3805712.3808631}, apply 5/20 user/item thresholds, and describe tracks by the released item embeddings.
MovieLens and Yambda have long histories, and Yambda is also repeat-heavy: 29\% of its test targets are among the user's last 100 items.

\begin{table}[t]
  \setlength{\abovecaptionskip}{6pt}
  \caption{Statistics of the datasets after preprocessing.}
  \label{tab:dataset_stats}
  \resizebox{1.0\columnwidth}{!}{%
      \centering
      \begin{tabular}{lrrrrrr}
      \toprule
          \textbf{Dataset} &
          \textbf{\#Users} &
          \textbf{\#Items} &
          \textbf{\#Interact.} &
          \textbf{Avg. Len.} &
          \textbf{Median Len.} &
          \textbf{Density (\%)} \\
      \midrule
          Beauty~\cite{mcauley2015imagebased} & 22,363 & 12,101 & 198,502 & 8.88 & 6.0 & 0.073 \\
          Sports~\cite{mcauley2015imagebased} & 35,598 & 18,357 & 296,337 & 8.32 & 6.0 & 0.045 \\
          Toys~\cite{mcauley2015imagebased} & 19,412 & 11,924 & 167,597 & 8.63 & 6.0 & 0.072 \\
          BPC-23~\cite{hou2024bridging} & 729,576 & 207,649 & 6,624,441 & 9.08 & 7.0 & 0.004 \\
          Books-23~\cite{hou2024bridging} & 776,370 & 495,063 & 9,488,297 & 12.22 & 7.0 & 0.002 \\
          MovieLens~\cite{harper2015movielens} & 110,456 & 10,870 & 16,193,689 & 146.61 & 69.0 & 1.349 \\
          Yambda~\cite{yambda} & 87,806 & 79,197 & 12,818,976 & 145.99 & 173.0 & 0.184 \\
      \bottomrule
      \end{tabular}
}
\end{table}

\subsubsection{Evaluation Protocol}
\label{sec:metrics}

We use a global temporal split~\cite{gusak2025time,10.1145/3805712.3808631}, which, unlike leave-one-out, restricts training to interactions that precede the evaluation periods.
The latest 10\% of interactions by timestamp define the test period and the latest 10\% of the rest the validation period (on Yambda, the last day and the day before).
Each user's last interaction in a held-out period is the target.
The input history includes earlier interactions both before and within that period~\cite{gusak2025time}.
Items the user has interacted with are not filtered from recommendations: repeats are legitimate targets, and in production such filtering typically follows retrieval.
We report NDCG@$k$ (N@$k$) and Recall@$k$ (R@$k$) over the full training catalog with beam width 100 for $k\in\{10,50,100\}$, and up to $k=1000$ in the candidate-budget study; main results are means and standard deviations over three seeds~\cite{gusak2026overconfidence}.

Efficiency is measured on one H100 GPU.
Our six models use a production-style pipeline with bf16 weights, compilation, CUDA graphs and a beam-shared history cache (Appendix~\ref{app:impl}).
Separately, since external baselines have different retrieval implementations, Table~\ref{tab:baselines} times them and \deco in a common eager PyTorch~\cite{paszke2019pytorch} execution setup, each through its native retrieval path.
Latency is the GPU time for prefill and beam search per batch; throughput is batch size divided by median latency.
We report throughput at the peak of a batch-size sweep, under a 50\,ms 95th-percentile (p95) latency budget, and at batch 128, the largest where all full-backbone baselines meet that budget, near their peak (Appendix~\ref{app:impl}).

\begin{table}[t]
  \setlength{\abovecaptionskip}{6pt}
  \caption{Accuracy (mean $\pm$ SD over 3 seeds; bold/underlined: best/second-best mean per dataset) and generation speed. \flatgr and \flatgrmatch are baselines (grey). Speedup over \flatgr at batch 128 / at peak throughput. \mkboth: mean exceeds both its teacher and \flatgr; \mkteach: exceeds its teacher only.}
  \label{tab:main}
  \small
  \setlength{\tabcolsep}{3pt}
  \begin{adjustbox}{max width=\columnwidth}
  \begin{tabular}{cl >{\hspace{7pt}}r >{\hspace{7pt}}r >{\hspace{7pt}}r >{\raggedleft\arraybackslash}p{28pt} >{\raggedleft\arraybackslash}p{28pt} r}
    \toprule
    & & \multicolumn{3}{c}{\textbf{Accuracy}} & \multicolumn{2}{c}{\textbf{Speed, users/s}} & \textbf{Speedup} \\
    \cmidrule(lr){3-5} \cmidrule(lr){6-7} \cmidrule(lr){8-8}
    & \textbf{Model} & \multicolumn{1}{c}{NDCG@10} & \multicolumn{1}{c}{Recall@50} & \multicolumn{1}{c}{Recall@100} & b128 & peak & b128\,/\,peak \\
    \midrule
     & \textcolor{basegray}{\flatgr} & 0.0221\pmsd{.0018} & 0.0960\pmsd{.0154} & 0.1319\pmsd{.0145} & 6,454 & 7,083 & \refslash \\
     & \textcolor{basegray}{\flatgrmatch} & 0.0216\pmsd{.0027} & 0.0856\pmsd{.0129} & 0.1178\pmsd{.0163} & 7,922 & 8,558 & \xs1.2\,/\,\xs1.2 \\
     & \decoflat & \llap{\mkboth\,}\underline{0.0308}\pmsd{.0009} & \llap{\mkboth\,}\underline{0.1249}\pmsd{.0002} & \llap{\mkboth\,}\underline{0.1671}\pmsd{.0036} & 22,362 & 29,737 & \xs3.5\,/\,\xs4.2 \\
    \addlinespace[2pt]
     & \fusedgr & 0.0300\pmsd{.0010} & 0.1147\pmsd{.0029} & 0.1526\pmsd{.0053} & 7,876 & 8,652 & \xs1.2\,/\,\xs1.2 \\
     & \decoce & 0.0287\pmsd{.0035} & 0.1081\pmsd{.0027} & 0.1439\pmsd{.0056} & 39,938 & 59,634 & \xs6.2\,/\,\xs8.4 \\
    \multirow{-6}{*}{\rotatebox[origin=c]{90}{Toys}} & \cellcolor{decobg}\deco & \cellcolor{decobg}\llap{\mkboth\,}\textbf{0.0352}\pmsd{.0009} & \cellcolor{decobg}\llap{\mkboth\,}\textbf{0.1296}\pmsd{.0058} & \cellcolor{decobg}\llap{\mkboth\,}\textbf{0.1718}\pmsd{.0049} & \cellcolor{decobg}39,853 & \cellcolor{decobg}59,451 & \cellcolor{decobg}\xs6.2\,/\,\xs8.4 \\
    \midrule
     & \textcolor{basegray}{\flatgr} & 0.0138\pmsd{.0024} & 0.0747\pmsd{.0088} & 0.1087\pmsd{.0139} & 6,458 & 6,993 & \refslash \\
     & \textcolor{basegray}{\flatgrmatch} & 0.0143\pmsd{.0030} & 0.0671\pmsd{.0075} & 0.1005\pmsd{.0070} & 7,917 & 8,889 & \xs1.2\,/\,\xs1.3 \\
     & \decoflat & \llap{\mkboth\,}0.0184\pmsd{.0018} & \llap{\mkboth\,}\textbf{0.0917}\pmsd{.0035} & \llap{\mkboth\,}\underline{0.1307}\pmsd{.0069} & 22,418 & 29,234 & \xs3.5\,/\,\xs4.2 \\
    \addlinespace[2pt]
     & \fusedgr & \underline{0.0190}\pmsd{.0010} & 0.0841\pmsd{.0051} & 0.1217\pmsd{.0028} & 7,872 & 8,531 & \xs1.2\,/\,\xs1.2 \\
     & \decoce & 0.0173\pmsd{.0006} & 0.0783\pmsd{.0035} & 0.1166\pmsd{.0056} & 39,924 & 59,682 & \xs6.2\,/\,\xs8.5 \\
    \multirow{-6}{*}{\rotatebox[origin=c]{90}{Beauty}} & \cellcolor{decobg}\deco & \cellcolor{decobg}\llap{\mkboth\,}\textbf{0.0200}\pmsd{.0004} & \cellcolor{decobg}\llap{\mkboth\,}\underline{0.0912}\pmsd{.0052} & \cellcolor{decobg}\llap{\mkboth\,}\textbf{0.1319}\pmsd{.0066} & \cellcolor{decobg}39,895 & \cellcolor{decobg}59,629 & \cellcolor{decobg}\xs6.2\,/\,\xs8.5 \\
    \midrule
     & \textcolor{basegray}{\flatgr} & 0.0132\pmsd{.0005} & 0.0713\pmsd{.0031} & 0.1053\pmsd{.0036} & 6,392 & 7,004 & \refslash \\
     & \textcolor{basegray}{\flatgrmatch} & 0.0132\pmsd{.0005} & 0.0676\pmsd{.0022} & 0.1027\pmsd{.0038} & 7,814 & 8,592 & \xs1.2\,/\,\xs1.2 \\
     & \decoflat & \llap{\mkboth\,}0.0160\pmsd{.0009} & \llap{\mkboth\,}\underline{0.0786}\pmsd{.0019} & \llap{\mkboth\,}\underline{0.1170}\pmsd{.0034} & 21,953 & 29,796 & \xs3.4\,/\,\xs4.3 \\
    \addlinespace[2pt]
     & \fusedgr & 0.0160\pmsd{.0003} & 0.0774\pmsd{.0004} & 0.1134\pmsd{.0008} & 7,765 & 8,547 & \xs1.2\,/\,\xs1.2 \\
     & \decoce & \underline{0.0164}\pmsd{.0004} & 0.0781\pmsd{.0040} & 0.1164\pmsd{.0035} & 38,878 & 59,632 & \xs6.1\,/\,\xs8.5 \\
    \multirow{-6}{*}{\rotatebox[origin=c]{90}{Sports}} & \cellcolor{decobg}\deco & \cellcolor{decobg}\llap{\mkboth\,}\textbf{0.0191}\pmsd{.0007} & \cellcolor{decobg}\llap{\mkboth\,}\textbf{0.0900}\pmsd{.0013} & \cellcolor{decobg}\llap{\mkboth\,}\textbf{0.1301}\pmsd{.0023} & \cellcolor{decobg}37,920 & \cellcolor{decobg}59,473 & \cellcolor{decobg}\xs5.9\,/\,\xs8.5 \\
    \midrule
     & \textcolor{basegray}{\flatgr} & 0.0851\pmsd{.0015} & 0.2984\pmsd{.0047} & 0.4031\pmsd{.0077} & 3,069 & 3,165 & \refslash \\
     & \textcolor{basegray}{\flatgrmatch} & 0.0680\pmsd{.0007} & 0.2602\pmsd{.0011} & 0.3599\pmsd{.0032} & 6,154 & 6,562 & \xs2.0\,/\,\xs2.1 \\
     & \decoflat & \llap{\mkboth\,}0.0856\pmsd{.0017} & \llap{\mkboth\,}0.3051\pmsd{.0014} & \llap{\mkboth\,}0.4122\pmsd{.0027} & 5,981 & 6,269 & \xs1.9\,/\,\xs2.0 \\
    \addlinespace[2pt]
     & \fusedgr & \underline{0.0887}\pmsd{.0016} & \underline{0.3083}\pmsd{.0037} & \underline{0.4134}\pmsd{.0054} & 6,113 & 6,604 & \xs2.0\,/\,\xs2.1 \\
     & \decoce & 0.0808\pmsd{.0013} & 0.2965\pmsd{.0019} & 0.3967\pmsd{.0005} & 20,090 & 24,622 & \xs6.5\,/\,\xs7.8 \\
    \multirow{-6}{*}{\rotatebox[origin=c]{90}{MovieLens}} & \cellcolor{decobg}\deco & \cellcolor{decobg}\llap{\mkboth\,}\textbf{0.0898}\pmsd{.0019} & \cellcolor{decobg}\llap{\mkboth\,}\textbf{0.3157}\pmsd{.0010} & \cellcolor{decobg}\llap{\mkboth\,}\textbf{0.4220}\pmsd{.0016} & \cellcolor{decobg}20,080 & \cellcolor{decobg}24,745 & \cellcolor{decobg}\xs6.5\,/\,\xs7.8 \\
    \midrule
     & \textcolor{basegray}{\flatgr} & \underline{0.0136}\pmsd{.0001} & \underline{0.0648}\pmsd{.0002} & \underline{0.0907}\pmsd{.0008} & 6,410 & 6,958 & \refslash \\
     & \textcolor{basegray}{\flatgrmatch} & 0.0128\pmsd{.0002} & 0.0624\pmsd{.0012} & 0.0879\pmsd{.0006} & 7,854 & 8,577 & \xs1.2\,/\,\xs1.2 \\
     & \decoflat & \llap{\mkboth\,}\textbf{0.0144}\pmsd{.0001} & \llap{\mkboth\,}\textbf{0.0675}\pmsd{.0005} & \llap{\mkboth\,}\textbf{0.0946}\pmsd{.0008} & 21,938 & 29,350 & \xs3.4\,/\,\xs4.2 \\
    \addlinespace[2pt]
     & \fusedgr & 0.0129\pmsd{.0001} & 0.0623\pmsd{.0004} & 0.0886\pmsd{.0003} & 7,810 & 8,497 & \xs1.2\,/\,\xs1.2 \\
     & \decoce & 0.0127\pmsd{.0001} & 0.0636\pmsd{.0008} & 0.0894\pmsd{.0007} & 38,393 & 57,533 & \xs6.0\,/\,\xs8.3 \\
    \multirow{-6}{*}{\rotatebox[origin=c]{90}{BPC-23}} & \cellcolor{decobg}\deco & \cellcolor{decobg}\llap{\mkteach\,}0.0132\pmsd{.0003} & \cellcolor{decobg}\llap{\mkteach\,}0.0635\pmsd{.0002} & \cellcolor{decobg}\llap{\mkteach\,}0.0905\pmsd{.0005} & \cellcolor{decobg}38,406 & \cellcolor{decobg}57,576 & \cellcolor{decobg}\xs6.0\,/\,\xs8.3 \\
    \midrule
     & \textcolor{basegray}{\flatgr} & 0.0193\pmsd{.0003} & 0.0739\pmsd{.0006} & \underline{0.0985}\pmsd{.0004} & 6,337 & 6,983 & \refslash \\
     & \textcolor{basegray}{\flatgrmatch} & 0.0165\pmsd{.0003} & 0.0671\pmsd{.0002} & 0.0905\pmsd{.0009} & 7,718 & 8,447 & \xs1.2\,/\,\xs1.2 \\
     & \decoflat & 0.0193\pmsd{.0002} & \llap{\mkboth\,}\underline{0.0747}\pmsd{.0004} & \llap{\mkboth\,}\textbf{0.1002}\pmsd{.0008} & 21,240 & 28,857 & \xs3.4\,/\,\xs4.1 \\
    \addlinespace[2pt]
     & \fusedgr & \underline{0.0196}\pmsd{.0004} & 0.0736\pmsd{.0008} & 0.0973\pmsd{.0010} & 7,682 & 8,750 & \xs1.2\,/\,\xs1.3 \\
     & \decoce & 0.0178\pmsd{.0001} & 0.0670\pmsd{.0004} & 0.0900\pmsd{.0007} & 35,885 & 55,910 & \xs5.7\,/\,\xs8.0 \\
    \multirow{-6}{*}{\rotatebox[origin=c]{90}{Books-23}} & \cellcolor{decobg}\deco & \cellcolor{decobg}\llap{\mkboth\,}\textbf{0.0197}\pmsd{.0005} & \cellcolor{decobg}\llap{\mkboth\,}\textbf{0.0753}\pmsd{.0010} & \cellcolor{decobg}\llap{\mkboth\,}\textbf{0.1002}\pmsd{.0010} & \cellcolor{decobg}35,926 & \cellcolor{decobg}55,848 & \cellcolor{decobg}\xs5.7\,/\,\xs8.0 \\
    \midrule
     & \textcolor{basegray}{\flatgr} & \underline{0.1746}\pmsd{.0008} & 0.3517\pmsd{.0010} & 0.4144\pmsd{.0022} & 2,943 & 3,023 & \refslash \\
     & \textcolor{basegray}{\flatgrmatch} & 0.0995\pmsd{.0013} & 0.2553\pmsd{.0028} & 0.3116\pmsd{.0036} & 5,827 & 6,370 & \xs2.0\,/\,\xs2.1 \\
     & \decoflat & \llap{\mkboth\,}\textbf{0.1779}\pmsd{.0004} & \llap{\mkboth\,}\textbf{0.3588}\pmsd{.0010} & \llap{\mkboth\,}\textbf{0.4229}\pmsd{.0015} & 5,518 & 5,812 & \xs1.9\,/\,\xs1.9 \\
    \addlinespace[2pt]
     & \fusedgr & 0.1674\pmsd{.0008} & 0.3478\pmsd{.0004} & 0.4125\pmsd{.0025} & 5,688 & 6,151 & \xs1.9\,/\,\xs2.0 \\
     & \decoce & 0.1647\pmsd{.0007} & 0.3466\pmsd{.0010} & 0.4099\pmsd{.0016} & 16,073 & 19,782 & \xs5.5\,/\,\xs6.5 \\
    \multirow{-6}{*}{\rotatebox[origin=c]{90}{Yambda}} & \cellcolor{decobg}\deco & \cellcolor{decobg}\llap{\mkteach\,}0.1712\pmsd{.0006} & \cellcolor{decobg}\llap{\mkboth\,}\underline{0.3544}\pmsd{.0006} & \cellcolor{decobg}\llap{\mkboth\,}\underline{0.4209}\pmsd{.0007} & \cellcolor{decobg}16,069 & \cellcolor{decobg}19,890 & \cellcolor{decobg}\xs5.5\,/\,\xs6.6 \\
    \bottomrule
  \end{tabular}
  \end{adjustbox}
\end{table}

\subsubsection{Models and Baselines}
\label{sec:models}

Six models share the backbone family, the SIDs and the training budget: \flatgr, a decoder-only TIGER-style model~\cite{rajput2023tiger}; \flatgrmatch, which truncates the flat history to $T/K$ items to read as many tokens as the fused models; \fusedgr, which combines each history item's code embeddings into a single input token; \deco, with delegated decoding and distillation from \fusedgr; \decoflat, its flat-history variant distilled from \flatgr; and \decoce, trained without a teacher using cross-entropy alone.

On Sports and BPC-23, a small and a large catalog, we also compare with SID-based generative recommenders tuned under our protocol and trained from scratch: TIGER~\cite{rajput2023tiger} with a KV cache, LLaDA-Rec~\cite{shi2025lladarec}, RPG~\cite{hou2025rpg} with both graph-based candidate search and full-catalog scoring, and NEZHA~\cite{wang2026nezha}.
We further include SASRec$^{+}$ (SASRec~\cite{kang2018self} with full cross-entropy), a state-of-the-art sequential recommender~\cite{klenitskiy2023turning,petrov2023gsasrec,mezentsev2024scalable} that scores the whole catalog instead of generating.
Finally, we compare with recent SID-based recommenders on the leave-one-out Beauty benchmark, using the results reported in the literature.

\subsubsection{Implementation Details}
\label{sec:implementation_details}

SIDs have $K=4$ levels of $V=2048$ codes, built by RQ-KMeans from multilingual E5-large text embeddings~\cite{wang2022text,wang2024multilingual} or from the provided 128-dimensional item embeddings on Yambda, with popularity-based collision resolution~\cite{zhou2025openonerec}.
The backbone is an 8-layer Qwen2-style decoder~\cite{yang2024qwen2} of width 640 with 10 heads, and the light decoder has 2 layers of width 256 with 4 heads.
The fuser embeds codes and mixes them with a two-layer MLP.
We keep input and output code tables untied for flexibility: input representations can include extra item features without changing the output classes; tying gives small gains for most models but does not change the comparison (Appendix~\ref{app:tying}).
The history is truncated to the last $T=20$ items on the Amazon datasets and $T=100$ on MovieLens and Yambda.
We train with Adam~\cite{kingma2015adam} and a warmup-cosine schedule, tuning the learning rate on validation.
We select checkpoints by validation N@10 at beam 10 and test at beam 100.
On Yambda, all models use the copy pointer of \S\ref{sec:copy}.
For the LLM backbones, we adapt Qwen2.5~\cite{qwen2025qwen25} with LoRA~\cite{hu2022lora}; before recommendation training, bidirectional metadata--SID grounding aligns the item representations with the pretrained model~\cite{zheng2024lcrec,kong2025minionerec}, and LoRA is merged into the backbone weights before measuring inference latency and throughput.

\subsection{Results}
\label{sec:results}
\subsubsection{RQ1: Recommendation Quality}
\label{sec:rq1}
\deco achieves \emph{comparable or better accuracy} than both \flatgr and \fusedgr, with 5.5--6.5$\times$ and 2.8--5.1$\times$ their throughput at batch 128, respectively. Table~\ref{tab:main} separates the effects of history fusion and delegated decoding.

\paragraph{Fusion retains comparable quality.}
At equal item counts, \fusedgr is 1.2--2.0$\times$ faster than \flatgr at batch 128 and raises NDCG@10 by 21--38\% on Amazon-2014, while staying within about 5\% on the four larger datasets.
At equal token budgets and similar speed, it beats \flatgrmatch, whose flat history covers only $T/K$ items, by 30\% in NDCG@10 on MovieLens and 68\% on Yambda.
Fusion thus keeps more history per token, reducing history processing cost while retaining competitive accuracy.

\paragraph{Moving the search preserves accuracy.}
\deco exceeds \fusedgr in NDCG@10 by 5--19\% on Amazon-2014 and 0.5--2.3\% on the four larger datasets.
Compared with the standard generative baseline \flatgr, its NDCG@10 is higher on five datasets and less than 3\% lower on BPC-23 and Yambda.
\decoflat also matches or exceeds \flatgr on every dataset.
A single backbone read thus suffices to match a teacher that attends to the history at every code level.
\decoce tests this interface \emph{without teacher supervision}: it learns from the same observed labels as \fusedgr, yet stays within 10\% of it in NDCG@10 and 8\% in Recall@100, and surpasses it on Sports and in BPC-23 recall.
The compact interface thus supports accurate retrieval without teacher supervision. Distillation turns the remaining gap into a consistent gain (\S\ref{sec:rq5}). Since \deco and \decoce share the same inference path, teacher supervision improves quality without adding backbone calls or decoder work.

\paragraph{Concept decoding helps either input.}
Neither input dominates: \fusedgr leads on Amazon-2014 and MovieLens, \flatgr on BPC-23 and Yambda, and Books-23 is mixed.
\decoflat keeps \flatgr's flat input and \deco keeps \fusedgr's fused input, so both gains come without changing the history representation.
Speedups also compound across stages: on MovieLens, fusion makes generation 2.0$\times$ faster and concept decoding another 3.3$\times$, 6.5$\times$ overall.

\paragraph{External baselines.}
In Table~\ref{tab:baselines}, \deco leads in NDCG@10 on both datasets and Recall@100 on BPC-23; on Sports, only SASRec$^{+}$ has slightly higher Recall@100.
Only two baselines are faster: parallel RPG with full-catalog scoring on Sports (15\% lower NDCG@10) and SASRec$^{+}$, which does not generate.
On the larger BPC-23 catalog, \deco leads generative baselines in accuracy and speed and beats SASRec$^{+}$ by 27\% in NDCG@10 and 15\% in Recall@100.

\begin{table}[t]
  \setlength{\abovecaptionskip}{6pt}
  \caption{External baselines tuned within our models' budgets (at most 79M parameters) and selected by validation NDCG@10. Speed (eager): users/s at batch 128. TIGER uses a KV cache; RPG\textsubscript{graph} timing excludes graph construction.}
  \label{tab:baselines}
  \setlength{\tabcolsep}{3pt}
  \resizebox{\columnwidth}{!}{%
  \begin{tabular}{l rrr rrr}
    \toprule
    & \multicolumn{3}{c}{\textbf{Sports}} & \multicolumn{3}{c}{\textbf{BPC-23}} \\
    \cmidrule(lr){2-4} \cmidrule(lr){5-7}
    \textbf{Model} & \multicolumn{1}{c}{N@10} & \multicolumn{1}{c}{R@100} & users/s, b128 & \multicolumn{1}{c}{N@10} & \multicolumn{1}{c}{R@100} & users/s, b128 \\
    \midrule
    TIGER (KV) & 0.0126\pmsd{.0008} & 0.1099\pmsd{.0028} & 523 & 0.0110\pmsd{.0007} & 0.0804\pmsd{.0023} & 521 \\
    LLaDA-Rec & 0.0127\pmsd{.0003} & 0.0914\pmsd{.0011} & 46 & 0.0088\pmsd{.0003} & 0.0627\pmsd{.0009} & 46 \\
    RPG\textsubscript{graph} & 0.0161\pmsd{.0006} & 0.1002\pmsd{.0040} & 573 & 0.0086\pmsd{.0001} & 0.0444\pmsd{.0003} & 461 \\
    RPG\textsubscript{full} & 0.0163\pmsd{.0005} & 0.1027\pmsd{.0042} & 21,854 & 0.0099\pmsd{.0001} & 0.0594\pmsd{.0003} & 5,706 \\
    NEZHA & 0.0135\pmsd{.0008} & 0.1165\pmsd{.0019} & 2,923 & 0.0120\pmsd{.0006} & 0.0822\pmsd{.0020} & 2,864 \\
    SASRec$^{+}$ & 0.0184\pmsd{.0002} & 0.1332\pmsd{.0011} & 46,757 & 0.0104\pmsd{.0000} & 0.0787\pmsd{.0017} & 65,922 \\
    \midrule
    \rowcolor{decobg}
    \deco & 0.0191\pmsd{.0007} & 0.1301\pmsd{.0023} & 6,238 & 0.0132\pmsd{.0003} & 0.0905\pmsd{.0005} & 6,132 \\
    \bottomrule
  \end{tabular}}
\end{table}

\paragraph{Leave-one-out.}
While our primary evaluation uses a global temporal split to avoid temporal leakage, we also report leave-one-out results on Beauty (Table~\ref{tab:loo}) for comparison with published SID-based recommenders. \deco reaches 0.0481 NDCG@10 and 0.0839 Recall@10, retaining its advantage under this protocol.

\begin{table}[t]
\setlength{\belowcaptionskip}{6pt}
\caption{Comparison with SID-based generative recommenders on Beauty using the \textit{leave-one-out} split with results reported in the literature~\cite{cheng2026capsid,hou2025rpg,rajput2023tiger}. The setup follows \cite{petrov2023gsasrec}.}
\label{tab:loo}
\resizebox{1\columnwidth}{!}{
\begin{tabular}{lcccccc>{\columncolor{decobg}}c}
\toprule
Metric & TIGER~\cite{rajput2023tiger} & LC-Rec~\cite{zheng2024lcrec} & LETTER~\cite{wang2024letter} & ETEGRec~\cite{liu2025etegrec} & DIGER~\cite{fu2026differentiable} & RPG~\cite{hou2025rpg} & \deco \\
\midrule
N@10 & 0.0384 & 0.0397 & 0.0406 & 0.0415 & 0.0431 & \underline{0.0464} & \textbf{0.0481} \\
R@10 & 0.0648 & 0.0675 & 0.0708 & 0.0725 & 0.0752 & \underline{0.0809} & \textbf{0.0839} \\
\bottomrule
\end{tabular}%
}
\end{table}

\Needspace{4\baselineskip}
\subsubsection{RQ2: Serving Efficiency}
\label{sec:rq2}
\begin{figure}
  \centering
  \includegraphics[width=\columnwidth]{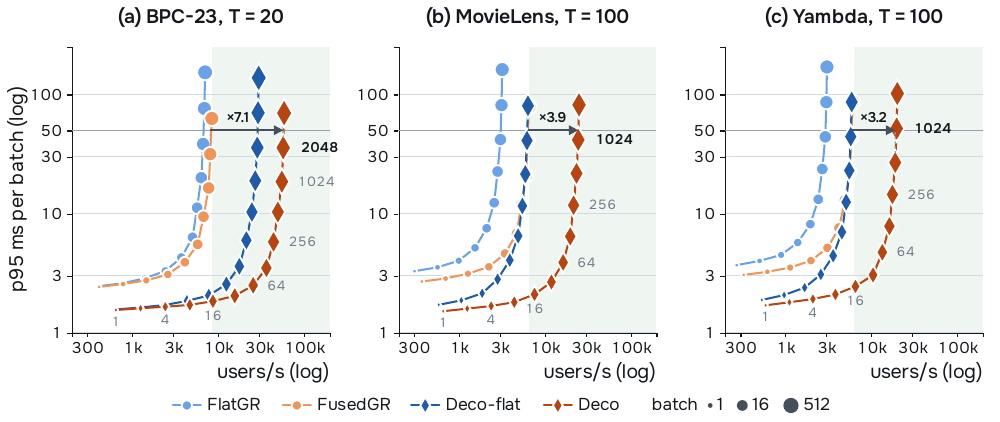}
  \caption{Serving curves (beam 100): p95 batch latency against users/s. Arrows at 50\,ms: within this p95 budget, \deco serves 7.1$\times$, 3.9$\times$ and 3.2$\times$ more users than \fusedgr. Shaded: throughput that no full-backbone model reaches.}
  \label{fig:serving}
\end{figure}

Fusion shortens prefill by compressing the history, and delegated decoding makes candidate expansion cheaper by assigning it to the light decoder.
On BPC-23 at batch 128, fusion cuts prefill from 4.4\,ms in \flatgr to 1.5\,ms in \fusedgr, but beam search still accounts for 91\% of generation time.
\deco then cuts search from 14.9 to 2.1\,ms, a 7$\times$ reduction.
Decoding therefore limits fusion's overall speedup, while delegation removes repeated backbone work from candidate expansion.

As the batch grows (Figure~\ref{fig:serving}), \deco's p95 latency stays under 8\,ms up to 128 users, whereas the full-backbone baselines saturate the GPU at a few hundred users per batch.
Under a 50\,ms p95 budget, \deco serves 7.1$\times$, 3.9$\times$ and 3.2$\times$ more users per second than \fusedgr on BPC-23, MovieLens and Yambda (6.5--8.6$\times$ more than \flatgr), and most of its curve lies at \emph{throughputs that no full-backbone model reaches}.
Thus, the accuracy gains in RQ1 come with greater serving capacity.

Our timings recompute the concept for every request, and on long histories, prefill becomes the main remaining cost.
Precomputing the 1\,KiB concept whenever a history changes (\S\ref{sec:deco}) leaves only light decoding for online requests.
With prefill excluded for both models, \deco decodes 4.8--7.7$\times$ faster than its teacher at batch 128 and up to 10.6$\times$ at peak.

\subsubsection{RQ3: Scaling the Generation Cost Factors}
\label{sec:rq3}
\begin{figure}
  \setlength{\abovecaptionskip}{6pt}
  \centering
  \includegraphics[width=\columnwidth]{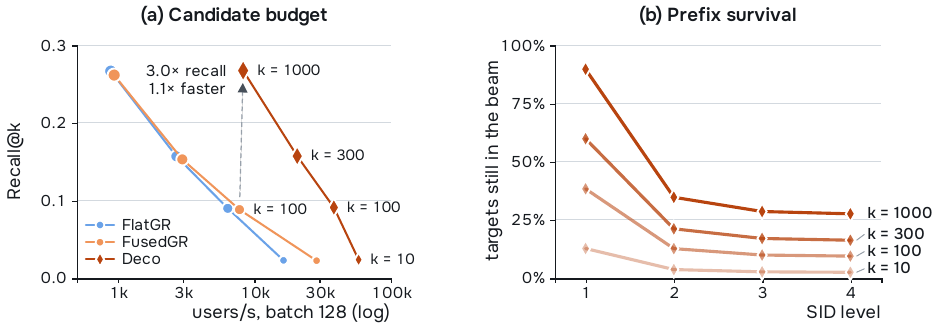}
  \caption{The candidate budget on BPC-23 (beam width $=k$). (a) Recall@$k$ against throughput: \deco at $k = 1000$ is faster than \fusedgr at $k = 100$, with 3$\times$ its recall. (b) Share of targets whose SID prefix survives in the \deco beam after each level (full-backbone models behave the same).}
  \label{fig:candidate-budget}
\end{figure}

\makeatletter\@secpenalty=10000\makeatother
\paragraph{Candidate budget.}
Returning the top $k$ items requires a beam of at least $k$, and widening it from 10 to 100 costs \deco 34\% of its throughput, against 73\% for \fusedgr and 61\% for \flatgr (Figure~\ref{fig:candidate-budget}a).
At $k=1000$, \deco still exceeds the throughput of \fusedgr at $k=100$ and recalls three times as many targets (0.267 vs.\ 0.088); against \fusedgr at $k=1000$, it is 8.8$\times$ faster at equal recall.
Wide beams pay off because most targets leave the beam at the first two SID levels (Fig.~\ref{fig:candidate-budget}b; at $k=100$, only 38\% survive the first and 13\% the second), and later levels cannot restore a pruned path.
Light decoding makes wider beams affordable, allowing \deco to preserve more early SID branches and return more candidates.
At fixed $k$, widening the beam beyond $k$ raises Recall@$k$ a little (Appendix~\ref{app:results}).
All models lose targets at similar rates across levels, suggesting that the concept does not make early decisions worse.

\makeatletter\@secpenalty=-300\makeatother

\begin{figure}
  \setlength{\abovecaptionskip}{6pt}
  \centering
  \includegraphics[width=\columnwidth]{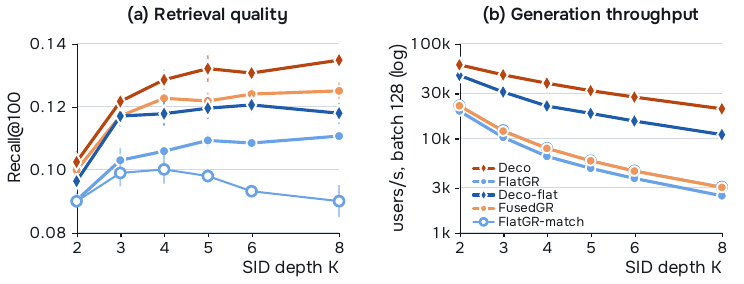}
  \caption{Increasing SID depth $K$ with nested codebooks on Sports ($V = 256$, $T=20$): (a) Recall@100; (b) throughput. \deco gains from finer identifiers without full-backbone decoding at each added level.}
  \label{fig:sid-depth}
\end{figure}

\paragraph{SID depth.}
Longer SIDs describe items more finely. As we vary the depth from $K=2$ to $8$ with nested codebooks of 256 codes, small enough that every added level still splits items (Figure~\ref{fig:sid-depth}), Recall@100 on Sports rises up to about $K=5$ and then levels off for all models except \flatgrmatch, which degrades beyond $K=4$ as its history shrinks to $20/K$ items (\S\ref{sec:rq1}).
With the same compact concept, \deco is the most accurate at every depth, so the concept need not grow with the identifier.
Each extra level lengthens both the input and the output of \flatgr, only the backbone decoding of \fusedgr, and only the light decoding of \deco.
From $K=2$ to $8$, \deco therefore slows down 2.9$\times$, against 7.3$\times$ for \fusedgr and 7.9$\times$ for \flatgr, and its speedup over its teacher grows from 2.7$\times$ to 6.8$\times$.
Extra levels never reach the backbone, which makes deeper SID hierarchies, such as PLUM's~\cite{he2025plum}, cheaper to serve.

\paragraph{History length.}
A longer history lengthens the prefill for both \deco and its teacher: from $T=20$ to $T=100$, \deco's speedup over the teacher falls from 4.7--5.1$\times$ to 2.8--3.3$\times$ but stays at 5.5--6.5$\times$ over \flatgr (Table~\ref{tab:main}). Excluding prefill, it decodes 6.7--7.7$\times$ faster than its teacher (4.8$\times$ with copying), while maintaining accuracy on both long-history datasets. Concept caching is therefore especially valuable for long histories.

\begin{figure}
  \setlength{\abovecaptionskip}{6pt}
  \centering
  \includegraphics[width=\columnwidth]{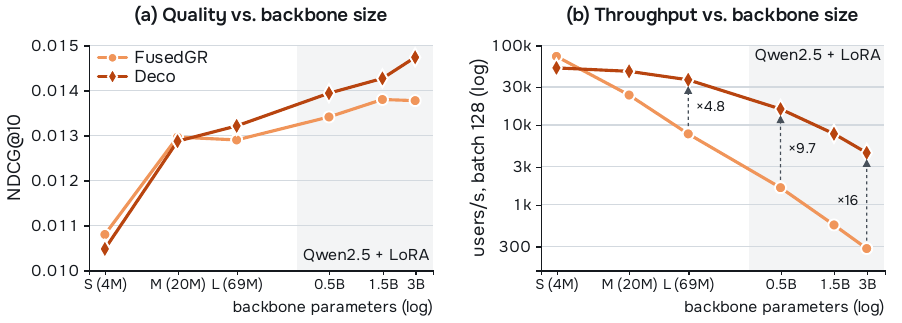}
  \caption{Scaling the backbone on BPC-23: S, M and L are trained from scratch; Qwen2.5 0.5B, 1.5B and 3B are fine-tuned with LoRA, which is merged into the weights for timing. Labels give the speedup over \fusedgr.}
  \label{fig:backbone-scaling}
\end{figure}

\paragraph{Backbone size and LLMs.}
On BPC-23, we scale the backbone from 4M to 69M parameters (S, M, L; trained from scratch) and on to Qwen2.5 with 0.5B, 1.5B and 3B parameters (Figure~\ref{fig:backbone-scaling}), keeping the same two-layer decoder.
From L onward, \deco stays ahead of its teacher, and the margin widens with scale (0.0147 vs.\ 0.0138 NDCG@10 at 3B), so a fixed-size concept does not cap what a larger backbone contributes.
From S to 3B, \fusedgr slows down 258$\times$ (73k to 281 users/s) as each beam step uses the larger backbone, whereas \deco, whose backbone runs once, slows down 12$\times$ (52k to 4.5k). All of \deco's added cost is in prefill: decoding stays at 58--59k users/s across backbones.
The speedup grows from 4.8$\times$ at L to 9.7$\times$, 14$\times$ and 16$\times$ at 0.5B, 1.5B and 3B.
\deco is slower only with backbone S, whose two layers of width 128 are lighter than the decoder, as Eqs.~\eqref{eq:cost-ar} and~\eqref{eq:cost-deco} predict.
Shrinking \fusedgr's backbone is no substitute for delegating the search: \fusedgr with the 20M backbone is both slower and less accurate than \deco with the 69M backbone L. With a 22$\times$ larger 1.5B backbone, \deco is \emph{as fast as the 69M \fusedgr and 11\% more accurate} in NDCG@10 (Figure~\ref{fig:backbone-tradeoff}, Appendix~\ref{app:results}).
\deco thus scales more efficiently and transfers to pretrained LLMs. Larger backbones improve retrieval quality without increasing the cost of candidate search.

\subsubsection{RQ4: Is a Compact Concept Enough?}
\label{sec:rq4}
\paragraph{Richer access to the history gives no consistent gain.}
In Table~\ref{tab:access}, we replace the concept with richer views of the history: all $T$ item states as a decoder prefix, the same states read by cross-attention in every decoder layer, as in the encoder--decoder access of TIGER and OneRec, and, for \decoflat, cross-attention to all $TK$ SID-token states.
NDCG@10 moves by less than 5\%, more often down than up, and Recall@100 gains at most about 3\%, while latency grows by up to 20\% and the per-user state from $O(1)$ to $O(T)$ or $O(TK)$.
On these low-repeat datasets, once the backbone has condensed the history into an expectation of the next item, the individual items add little that the decoder can use, so a decoder that never sees the history is both cheaper and accurate enough to serve.

\begin{table}[t]
  \setlength{\abovecaptionskip}{6pt}
  \caption{Decoder input: the concept $z$ (default, shaded), all item states as a prefix or via cross-attention (flat history: each item's last SID-token state), or all SID-token states. Input size: growth of decoder input and per-user state with $T$.}
  \label{tab:access}
  \small
  \setlength{\tabcolsep}{3pt}
  \setlength{\aboverulesep}{0pt}\setlength{\belowrulesep}{0pt}\renewcommand{\arraystretch}{1.2}  
  \begin{adjustbox}{max width=\columnwidth}
  \begin{tabular}{l l c cc c ccc}
    \toprule
    & & & \multicolumn{2}{c}{\textbf{Sports}} & & \multicolumn{3}{c}{\textbf{BPC-23}} \\
    \cmidrule(lr){4-5} \cmidrule(lr){7-9}
     & \textbf{Decoder input} & \textbf{Input size} & N@10 & R@100 & & N@10 & R@100 &  ms, b128\,{\scriptsize$\downarrow$} \\
    \midrule
     & \cellcolor{decobg}Last state & \cellcolor{decobg}O(1) & \cellcolor{decobg}0.0191\pmsd{.0007} & \cellcolor{decobg}0.1301\pmsd{.0023} & \cellcolor{decobg} & \cellcolor{decobg}0.0132\pmsd{.0003} & \cellcolor{decobg}0.0905\pmsd{.0005} & \cellcolor{decobg}3.32 \\
     & All item states, as prefix & O(T) & 0.0185\pmsd{.0004} & 0.1293\pmsd{.0016} &  & 0.0134\pmsd{.0003} & 0.0911\pmsd{.0007} & 3.40 \\
    \multirow{-3}{*}{\rotatebox[origin=c]{90}{\deco}} & All item states, cross-attn. & O(T) & 0.0186\pmsd{.0005} & 0.1285\pmsd{.0033} &  & 0.0132\pmsd{.0003} & 0.0913\pmsd{.0009} & 3.97 \\
    \midrule
     & \cellcolor{decobg}Last state & \cellcolor{decobg}O(1) & \cellcolor{decobg}0.0160\pmsd{.0009} & \cellcolor{decobg}0.1170\pmsd{.0034} & \cellcolor{decobg} & \cellcolor{decobg}0.0144\pmsd{.0001} & \cellcolor{decobg}0.0946\pmsd{.0008} & \cellcolor{decobg}5.85 \\
     & All item states, as prefix & O(T) & 0.0155\pmsd{.0017} & 0.1198\pmsd{.0052} &  & 0.0143\pmsd{.0001} & 0.0948\pmsd{.0005} & 5.91 \\
     & All item states, cross-attn. & O(T) & 0.0153\pmsd{.0006} & 0.1179\pmsd{.0062} &  & 0.0144\pmsd{.0002} & 0.0951\pmsd{.0013} & 6.50 \\
    \multirow{-4}{*}{\rotatebox[origin=c]{90}{\decoflat}} & All SID-token states, cross-attn. & O(TK) & 0.0154\pmsd{.0002} & 0.1207\pmsd{.0040} &  & 0.0146\pmsd{.0001} & 0.0962\pmsd{.0006} & 6.59 \\
    \bottomrule
  \end{tabular}
  \end{adjustbox}
\end{table}

\begin{figure}[t]
  \setlength{\abovecaptionskip}{6pt}
  \centering
  \includegraphics[width=1.0\columnwidth]{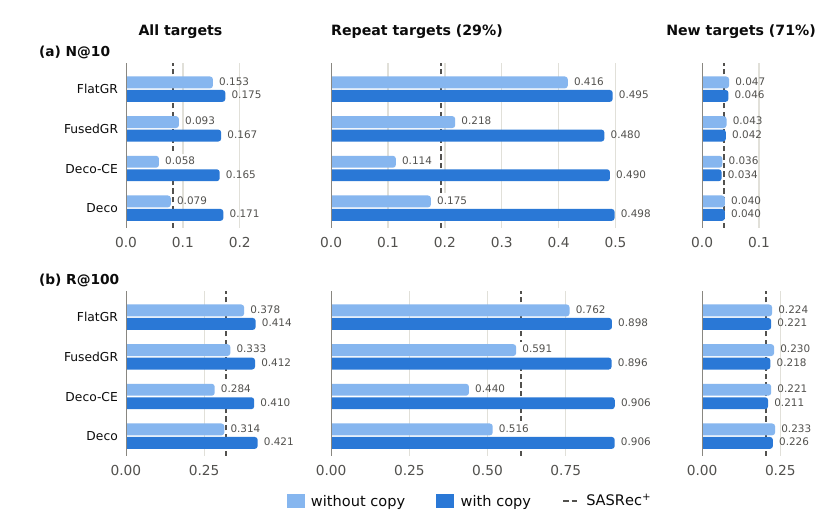}
  \caption{The copy pointer on Yambda: quality without (light bars) and with (dark bars) the pointer, on all, repeat (among the last 100 items) and new targets. Dashed line: SASRec$^{+}$.}
  \label{fig:yambda-copy}
\end{figure}

\paragraph{Repeat consumption.}
Without a pointer (light bars in Figure~\ref{fig:yambda-copy}), the fused models trail \flatgr on repeat-heavy Yambda by a wide margin: 0.093 NDCG@10 for \fusedgr and 0.079 for \deco against 0.153.
More than 90\% of the gap comes from repeat targets, while on new targets all four generators and SASRec$^{+}$ are close (0.036--0.047 NDCG@10).
The flat history exposes the exact codes of every past track for \flatgr's attention to copy; fused tokens mix them, and a compact concept encodes what kind of track comes next rather than which past track to replay.
The item-level SASRec$^{+}$ shows the same split: it matches every generator on new targets but falls far behind \flatgr on repeats, the source of most of \flatgr's lead, so on repeat-heavy data like Yambda, comparing SID generators with item-ID models largely measures copying ability.

With the pointer of \S\ref{sec:copy} (dark bars), every model reaches about 0.90 Recall@100 and 0.48--0.50 NDCG@10 on repeat targets, while metrics on new targets change by at most 0.012.
\deco's NDCG@10 more than doubles, from 0.079 to 0.171, and the four models end up within 0.011 Recall@100 of each other, with \deco the highest at 0.421.
The gate learns when to point without being told: it is computed from the history alone, yet its mean, 0.27, is close to the share of repeat targets, 0.29.
It averages 0.63 before repeat targets and 0.12 before new ones (AUC 0.90), and on repeat targets 86\% of the target's probability comes from the pointer; all full-history generators learn nearly the same gate.
The concept and pointer thus divide the work like semantic and episodic memory: one decides what kind of item comes next, the other retrieves the exact items the user returns to, and neither needs the backbone during search.\looseness=-1

\subsubsection{RQ5: Sizing and Training the Light Decoder}
\label{sec:rq5}
\makeatletter\@secpenalty=10000\makeatother
\paragraph{A small decoder suffices.}
Decoders with 1--4 layers of width 128--512 (3.5M to 21.6M parameters; Appendix~\ref{app:results}) stay within about 3\% of the default in NDCG@10 and Recall@100 on Sports and BPC-23, and a single layer cuts latency by 20--28\%: once the concept is given, spelling it out is a small job.
Our default is thus conservative; a smaller decoder offers a further latency--quality trade-off.

\begin{figure}
  \setlength{\abovecaptionskip}{6pt}
  \centering
\includegraphics[width=\columnwidth]{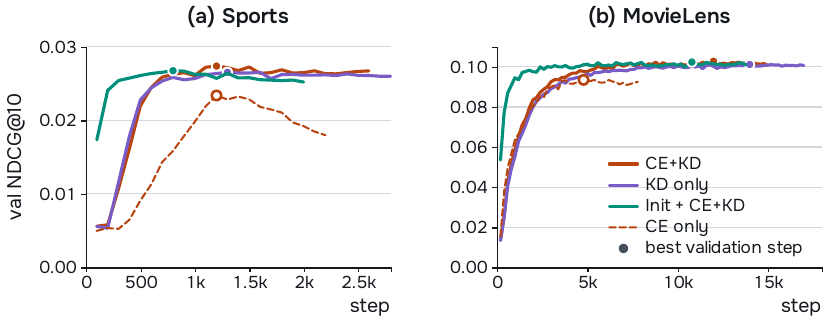}
  \caption{Validation dynamics of \deco: distillation improves peak quality, and teacher initialization speeds up early progress. CE+KD is \deco; CE only is \decoce; Init + CE+KD loads the teacher’s backbone and fuser.}
  \label{fig:val-dynamics}
\end{figure}

\makeatletter\@secpenalty=-300\makeatother
\paragraph{KD adds a consistent gain.}
Distillation raises N@10 over \decoce on every dataset (Table~\ref{tab:main}), and Figure~\ref{fig:val-dynamics} shows why.
With cross-entropy alone, the decoder overfits on Sports (validation N@10 peaks at 0.023 after 1.2k steps, then declines) and plateaus early at 0.094 on MovieLens.
With distillation, training is stable and reaches a higher level, 0.027 on Sports and 0.101--0.103 on MovieLens.
KD alone nearly matches CE+KD.
As teacher-informed label smoothing (Eq.~\eqref{eq:interp}), distillation should help most when labels are scarce; indeed, it adds 16--23\% N@10 on the small Amazon-2014 datasets but 4--11\% on the four larger ones.
We also train a full-backbone \fusedgr student with the same KD objective, but its accuracy gains are smaller and it remains less accurate than \deco (Appendix~\ref{app:results}). What matters for serving is that KD makes a far cheaper decoder at least as accurate as the model it replaces.

\paragraph{Teacher initialization is optional.}
A student initialized from the teacher's backbone and fuser reaches comparable quality in fewer steps (Figure~\ref{fig:val-dynamics}, green curves), which can shorten training; our main comparisons use independently initialized students.

\section{Conclusion}
\label{sec:conclusion}
Generative recommenders that run beam search in the backbone often spend most of their computation spelling identifiers rather than modeling user preferences.
\deco separates these tasks: the backbone reads the history once into a next-item concept and delegates the entire beam search to a light decoder.
Across seven datasets, it matches or exceeds its full-backbone teacher's accuracy and serves \mbox{3.2--7.3$\times$} more users under a 50\,ms p95 latency budget.
Taking the backbone out of candidate expansion makes generative retrieval affordable even with LLM-scale backbones and allows thousand-candidate pools, common before reranking in dense retrieval.
Refreshing the cached concept before requests arrive removes the backbone from online generation.
Together, these results show that the user-modeling backbone can scale \emph{without increasing the cost of candidate search}.

\bibliographystyle{ACM-Reference-Format}
\bibliography{bibliography}

@inproceedings{lin2025order,
title     = {Order-agnostic Identifier for Large Language Model-based Generative Recommendation},
author    = {Lin, Xinyu and Shi, Haihan and Wang, Wenjie and Feng, Fuli and Wang, Qifan and Ng, See-Kiong and Chua, Tat-Seng},
booktitle = {Proceedings of the 48th International ACM SIGIR Conference on Research and Development in Information Retrieval},
pages     = {1923--1933},
year      = {2025},
doi       = {10.1145/3726302.3730053}
}

@inproceedings{mezentsev2024scalable,
title     = {Scalable Cross-Entropy Loss for Sequential Recommendations with Large Item Catalogs},
author    = {Mezentsev, Gleb and Gusak, Danil and Oseledets, Ivan and Frolov, Evgeny},
booktitle = {Proceedings of the 18th ACM Conference on Recommender Systems},
pages     = {475--485},
year      = {2024},
doi       = {10.1145/3640457.3688140}
}

@inproceedings{gusak2025time,
  title={Time to Split: Exploring Data Splitting Strategies for Offline Evaluation of Sequential Recommenders},
  author={Gusak, Danil and Volodkevich, Anna and Klenitskiy, Anton and Vasilev, Alexey and Frolov, Evgeny},
  booktitle={Proceedings of the 19th ACM Conference on Recommender Systems},
  pages={874--883},
  year={2025}
}

@article{zhou2025openonerec,
    title   = {{OpenOneRec} Technical Report},
    author  = {Zhou, Guorui and Bao, Honghui and Huang, Jiaming and Deng, Jiaxin and Zhang,
  Jinghao and others},
    journal = {arXiv preprint arXiv:2512.24762},
    year    = {2025}
  }

@article{volodkevich2025autoregressive,
  author  = {Volodkevich, Anna and Gusak, Danil and Klenitskiy, Anton and Pembek, Anton and Vasilev, Alexey},
  title   = {Autoregressive generation strategies for {Top-K} sequential recommendations},
  journal = {User Modeling and User-Adapted Interaction},
  volume  = {35},
  number  = {3},
  articleno = {13},
  year    = {2025},
  doi     = {10.1007/s11257-025-09433-5}
}

@inproceedings{repeatconsumption2020doi,
  title = {Explainable Recommendation for Repeat Consumption},
  booktitle = {Proceedings of the 14th ACM Conference on Recommender Systems},
  year  = {2020},
  author = {Tsukuda, Kosetsu and Goto, Masataka},
  doi   = {10.1145/3383313.3412230}
}

@inproceedings{kang2018self,
  title={Self-Attentive Sequential Recommendation},
  author={Kang, Wang-Cheng and McAuley, Julian},
  booktitle={2018 IEEE International Conference on Data Mining (ICDM)},
  pages={197--206},
  year={2018},
  organization={IEEE}
}

@inproceedings{hou2025actionpiece,
  title={{ActionPiece}: Contextually Tokenizing Action Sequences for Generative Recommendation},
  author={Hou, Yupeng and Ni, Jianmo and He, Zhankui and Sachdeva, Noveen and Kang, Wang-Cheng and Chi, Ed H. and McAuley, Julian and Cheng, Derek Zhiyuan},
  booktitle={Proceedings of the 42nd International Conference on Machine Learning (ICML)},
  series={Proceedings of Machine Learning Research},
  volume={267},
  pages={24004--24024},
  year={2025}
}

@inproceedings{zhai2024actions,
  title={Actions Speak Louder than Words: Trillion-Parameter Sequential Transducers for Generative Recommendations},
  author={Zhai, Jiaqi and Liao, Lucy and Liu, Xing and Wang, Yueming and Li, Rui and Cao, Xuan and Gao, Leon and Gong, Zhaojie and Gu, Fangda and He, Jiayuan and Lu, Yinghai and Shi, Yu},
  booktitle={Proceedings of the 41st International Conference on Machine Learning (ICML)},
  series={Proceedings of Machine Learning Research},
  volume={235},
  pages={58484--58509},
  year={2024}
}

@inproceedings{petrov2023gsasrec,
  title={{gSASRec}: Reducing overconfidence in sequential recommendation trained with negative sampling},
  author={Petrov, Aleksandr Vladimirovich and Macdonald, Craig},
  booktitle={Proceedings of the 17th ACM Conference on Recommender Systems},
  pages={116--128},
  year={2023}
}

@inproceedings{klenitskiy2023turning,
  title={Turning dross into gold loss: is {BERT4Rec} really better than {SASRec}?},
  author={Klenitskiy, Anton and Vasilev, Alexey},
  booktitle={Proceedings of the 17th ACM Conference on Recommender Systems},
  pages={1120--1125},
  year={2023}
}

@article{deng2025onerec,
  title={OneRec: Unifying Retrieve and Rank with Generative Recommender and Iterative Preference Alignment},
  author={Deng, Jiaxin and Wang, Shiyao and Cai, Kuo and Ren, Lejian and Hu, Qigen and Ding, Weifeng and Luo, Qiang and Zhou, Guorui},
  journal={arXiv preprint arXiv:2502.18965},
  year={2025}
}

@inproceedings{kingma2015adam,
  title={Adam: A Method for Stochastic Optimization},
  author={Kingma, Diederik P. and Ba, Jimmy},
  booktitle={Proceedings of the 3rd International Conference on Learning Representations (ICLR)},
  year={2015}
}

@inproceedings{mcauley2015imagebased,
  author    = {McAuley, Julian and Targett, Christopher and Shi, Qinfeng and van den Hengel, Anton},
  title     = {Image-Based Recommendations on Styles and Substitutes},
  booktitle = {Proceedings of the 38th International ACM SIGIR Conference on Research and Development in Information Retrieval},
  pages     = {43--52},
  year      = {2015},
  doi       = {10.1145/2766462.2767755}
}

@inproceedings{tay2022transformer,
  title        = {Transformer Memory as a Differentiable Search Index},
  author       = {Tay, Yi and Tran, Vinh Q. and Dehghani, Mostafa and Ni, Jianmo and Bahri, Dara and Mehta, Harsh and Qin, Zhen and Hui, Kai and Zhao, Zhe and Gupta, Jai and Schuster, Tal and Cohen, William W. and Metzler, Donald},
  booktitle    = {Advances in Neural Information Processing Systems},
  year         = {2022}
}

@inproceedings{sun2023learning,
  title        = {Learning to Tokenize for Generative Retrieval},
  author       = {Sun, Weiwei and Yan, Lingyong and Chen, Zheng and Wang, Shuaiqiang and Zhu, Haichao and Ren, Pengjie and Chen, Zhumin and Yin, Dawei and de Rijke, Maarten and Ren, Zhaochun},
  booktitle    = {Advances in Neural Information Processing Systems},
  year         = {2023}
}

@article{zeghidour2022soundstream,
  title        = {{SoundStream}: An End-to-End Neural Audio Codec},
  author       = {Zeghidour, Neil and Luebs, Alejandro and Omran, Ahmed and Skoglund, Jan and Tagliasacchi, Marco},
  journal      = {IEEE/ACM Transactions on Audio, Speech, and Language Processing},
  volume       = {30},
  pages        = {495--507},
  year         = {2022},
  doi          = {10.1109/TASLP.2021.3129994}
}

@inproceedings{lee2022autoregressive,
  title        = {Autoregressive Image Generation Using Residual Quantization},
  author       = {Lee, Doyup and Kim, Chiheon and Kim, Saehoon and Cho, Minsu and Han, Wook-Shin},
  booktitle    = {Proceedings of the IEEE/CVF Conference on Computer Vision and Pattern Recognition},
  year         = {2022},
  doi          = {10.1109/CVPR52688.2022.01123}
}

@inproceedings{zhu2024cost,
  title        = {{CoST}: Contrastive Quantization Based Semantic Tokenization for Generative Recommendation},
  author       = {Zhu, Jieming and Jin, Mengqun and Liu, Qijiong and Qiu, Zexuan and Dong, Zhenhua and Li, Xiu},
  booktitle    = {Proceedings of the 18th ACM Conference on Recommender Systems},
  year         = {2024},
  doi          = {10.1145/3640457.3688178}
}

@inproceedings{liu2025understanding,
  title        = {Understanding Generative Recommendation with Semantic {IDs} from a Model-scaling View},
  author       = {Liu, Jingzhe and Collins, Liam and Tang, Jiliang and Zhao, Tong and Shah, Neil and Ju, Mingxuan},
  booktitle    = {Proceedings of the 32nd ACM SIGKDD Conference on Knowledge Discovery and Data Mining V.2},
  pages        = {3234--3245},
  year         = {2026},
  doi          = {10.1145/3770855.3817976}
}

@article{liang2026rethinking,
  title        = {Rethinking Generative Recommender Tokenizer: Recsys-Native Encoding and Semantic Quantization Beyond {LLMs}},
  author       = {Liang, Yu and Zhang, Zhongjin and Zhu, Yuxuan and Zhang, Kerui and Guo, Zhiluohan and Zhou, Wenhang and Yang, Zonqi and Wu, Kangle and Ni, Yabo and Zeng, Anxiang and Fu, Cong and Wang, Jianxin and Xia, Jiazhi},
  journal      = {arXiv preprint arXiv:2602.02338},
  year         = {2026}
}

@inproceedings{fu2026differentiable,
  title        = {Differentiable Semantic {ID} for Generative Recommendation},
  author       = {Fu, Junchen and Ge, Xuri and Karatzoglou, Alexandros and Arapakis, Ioannis and Verberne, Suzan and Jose, Joemon M. and Ren, Zhaochun},
  booktitle    = {Proceedings of the 49th International ACM SIGIR Conference on Research and Development in Information Retrieval},
  pages        = {369--379},
  year         = {2026},
  doi          = {10.1145/3805712.3809641}
}

@inproceedings{yambda,
  title         = {{Yambda-5B} -- A Large-Scale Multi-Modal Dataset for Ranking and Retrieval},
  author        = {Ploshkin, Alexander and Tytskiy, Vladislav and Pismenny, Alexey and Baikalov, Vladimir and Taychinov, Evgeny and Permiakov, Artem and Burlakov, Daniil and Krofto, Eugene},
  booktitle     = {Proceedings of the Nineteenth ACM Conference on Recommender Systems},
  pages         = {894--901},
  year          = {2025},
  doi           = {10.1145/3705328.3748163}
}

@article{hinton2015distilling,
  title={Distilling the knowledge in a neural network},
  author={Hinton, Geoffrey and Vinyals, Oriol and Dean, Jeff},
  journal={arXiv preprint arXiv:1503.02531},
  year={2015}
}

@article{wang2022text,
  title={Text embeddings by weakly-supervised contrastive pre-training},
  author={Wang, Liang and Yang, Nan and Huang, Xiaolong and Jiao, Binxing and Yang, Linjun and Jiang, Daxin and Majumder, Rangan and Wei, Furu},
  journal={arXiv preprint arXiv:2212.03533},
  year={2022}
}

@inproceedings{hou2024bridging,
  title={Bridging Language and Items for Retrieval and Recommendation: Benchmarking {LLM}s as Semantic Encoders},
  author={Hou, Yupeng and Li, Jiacheng and Fu, Xiangjun and He, Zhankui and Yan, An and Chen, Xiusi and McAuley, Julian},
  booktitle={Proceedings of the 64th Annual Meeting of the Association for Computational Linguistics (Volume 1: Long Papers)},
  pages={3251--3265},
  year={2026},
  doi={10.18653/v1/2026.acl-long.147},
  url={https://aclanthology.org/2026.acl-long.147/}
}

@inproceedings{lin2025efficient,
  title={Efficient inference for large language model-based generative recommendation},
  author={Lin, Xinyu and Yang, Chaoqun and Wang, Wenjie and Li, Yongqi and Du, Cunxiao and Feng, Fuli and Ng, See-Kiong and Chua, Tat-Seng},
  booktitle={Proceedings of the 13th International Conference on Learning Representations (ICLR)},
  year={2025}
}

@inproceedings{furlanello2018born,
  title={Born again neural networks},
  author={Furlanello, Tommaso and Lipton, Zachary and Tschannen, Michael and Itti, Laurent and Anandkumar, Anima},
  booktitle={Proceedings of the 35th International Conference on Machine Learning (ICML)},
  pages={1607--1616},
  year={2018}
}

@article{yang2024qwen2,
  title={Qwen2 technical report},
  author={Yang, An and Yang, Baosong and Hui, Binyuan and Zheng, Bo and Yu, Bowen and Zhou, Chang and Li, Chengpeng and Li, Chengyuan and Liu, Dayiheng and Huang, Fei and others},
  journal={arXiv preprint arXiv:2407.10671},
  year={2024}
}

@inproceedings{gusak2026overconfidence,
  title={More Test Users, More Overconfidence: Seed Variance in Sequential Recommender Tests},
  author={Gusak, Danil and Volodkevich, Anna and Frolov, Evgeny},
  booktitle={Proceedings of the 20th ACM Conference on Recommender Systems},
  doi={10.1145/3773078.3841281},
  year={2026}
}

@inproceedings{10.1145/3805712.3808631,
author = {Volodkevich, Anna and Anikin, Dmitry and Gusak, Danil and Klenitskiy, Anton and Frolov, Evgeny and Vasilev, Alexey},
title = {SplitLight: An Exploratory Toolkit for Recommender Systems Datasets and Splits},
year = {2026},
doi = {10.1145/3805712.3808631},
booktitle = {Proceedings of the 49th International ACM SIGIR Conference on Research and Development in Information Retrieval}
}

@inproceedings{10.1145/3805712.3808437,
author = {Zou, Yanyan and Qi, Junbo and Huang, Lunsong and Li, Yu and Xu, Kewei and Gao, Jiahao and Zhao, Binglei and Yang, Xuanhua and Xu, Sulong and Li, Shengjie},
title = {GenRec: A Preference-Oriented Generative Framework for Large-Scale Recommendation},
year = {2026},
doi = {10.1145/3805712.3808437},
booktitle = {Proceedings of the 49th International ACM SIGIR Conference on Research and Development in Information Retrieval}
}

@article{harper2015movielens,
  author    = {Harper, F. Maxwell and Konstan, Joseph A.},
  title     = {The {MovieLens} Datasets: History and Context},
  journal   = {ACM Transactions on Interactive Intelligent Systems},
  year      = {2015},
  publisher = {Association for Computing Machinery},
  doi       = {10.1145/2827872},
  url       = {https://doi.org/10.1145/2827872}
}

@inproceedings{rajput2023tiger,
  title     = {Recommender Systems with Generative Retrieval},
  author    = {Rajput, Shashank and Mehta, Nikhil and Singh, Anima and Keshavan, Raghunandan H. and Vu, Trung and Heldt, Lukasz and Hong, Lichan and Tay, Yi and Tran, Vinh Q. and Samost, Jonah and Kula, Maciej and Chi, Ed H. and Sathiamoorthy, Maheswaran},
  booktitle = {Advances in Neural Information Processing Systems 36 (NeurIPS 2023)},
  pages     = {10299--10315},
  publisher = {Curran Associates, Inc.},
  year      = {2023}
}

@inproceedings{zheng2024lcrec,
  title     = {Adapting Large Language Models by Integrating Collaborative Semantics for Recommendation},
  author    = {Zheng, Bowen and Hou, Yupeng and Lu, Hongyu and Chen, Yu and Zhao, Wayne Xin and Chen, Ming and Wen, Ji-Rong},
  booktitle = {2024 IEEE 40th International Conference on Data Engineering (ICDE)},
  pages     = {1435--1448},
  publisher = {IEEE},
  year      = {2024},
  doi       = {10.1109/ICDE60146.2024.00118}
}

@inproceedings{wang2024letter,
  title     = {Learnable Item Tokenization for Generative Recommendation},
  author    = {Wang, Wenjie and Bao, Honghui and Lin, Xinyu and Zhang, Jizhi and Li, Yongqi and Feng, Fuli and Ng, See-Kiong and Chua, Tat-Seng},
  booktitle = {Proceedings of the 33rd ACM International Conference on Information and Knowledge Management},
  series    = {CIKM '24},
  pages     = {2400--2409},
  publisher = {Association for Computing Machinery},
  address   = {New York, NY, USA},
  year      = {2024},
  doi       = {10.1145/3627673.3679569}
}

@article{yang2025liger,
  title     = {Unifying Generative and Dense Retrieval for Sequential Recommendation},
  author    = {Yang, Liu and Paischer, Fabian and Hassani, Kaveh and Li, Jiacheng and Shao, Shuai and Li, Zhang Gabriel and He, Yun and Feng, Xue and Noorshams, Nima and Park, Sem and Long, Bo and Nowak, Robert D. and Gao, Xiaoli and Eghbalzadeh, Hamid},
  journal   = {Transactions on Machine Learning Research},
  year      = {2025},
  url       = {https://openreview.net/forum?id=jxdnFIsjCb}
}

@inproceedings{liu2025etegrec,
  title     = {Generative Recommender with End-to-End Learnable Item Tokenization},
  author    = {Liu, Enze and Zheng, Bowen and Ling, Cheng and Hu, Lantao and Li, Han and Zhao, Wayne Xin},
  booktitle = {Proceedings of the 48th International ACM SIGIR Conference on Research and Development in Information Retrieval},
  series    = {SIGIR '25},
  pages     = {729--739},
  publisher = {Association for Computing Machinery},
  address   = {New York, NY, USA},
  year      = {2025},
  doi       = {10.1145/3726302.3729989}
}

@inproceedings{hou2025rpg,
  title     = {Generating Long Semantic IDs in Parallel for Recommendation},
  author    = {Hou, Yupeng and Li, Jiacheng and Shin, Ashley and Jeon, Jinsung and Santhanam, Abhishek and Shao, Wei and Hassani, Kaveh and Yao, Ning and McAuley, Julian},
  booktitle = {Proceedings of the 31st ACM SIGKDD Conference on Knowledge Discovery and Data Mining V.2},
  series    = {KDD '25},
  pages     = {956--966},
  publisher = {Association for Computing Machinery},
  address   = {New York, NY, USA},
  year      = {2025},
  doi       = {10.1145/3711896.3736979}
}

@inproceedings{paszke2019pytorch,
  title     = {{PyTorch}: An Imperative Style, High-Performance Deep Learning Library},
  author    = {Paszke, Adam and Gross, Sam and Massa, Francisco and Lerer, Adam and Bradbury, James and Chanan, Gregory and Killeen, Trevor and Lin, Zeming and Gimelshein, Natalia and Antiga, Luca and Desmaison, Alban and K{\"o}pf, Andreas and Yang, Edward and DeVito, Zachary and Raison, Martin and Tejani, Alykhan and Chilamkurthy, Sasank and Steiner, Benoit and Fang, Lu and Bai, Junjie and Chintala, Soumith},
  booktitle = {Advances in Neural Information Processing Systems 32 (NeurIPS)},
  year      = {2019}
}

@article{juravsky2024hydragen,
  title   = {Hydragen: High-Throughput {LLM} Inference with Shared Prefixes},
  author  = {Juravsky, Jordan and Brown, Bradley and Ehrlich, Ryan and Fu, Daniel Y. and R{\'e}, Christopher and Mirhoseini, Azalia},
  journal = {arXiv preprint arXiv:2402.05099},
  year    = {2024}
}

@inproceedings{ye2025flashinfer,
  title     = {{FlashInfer}: Efficient and Customizable Attention Engine for {LLM} Inference Serving},
  author    = {Ye, Zihao and Chen, Lequn and Lai, Ruihang and Lin, Wuwei and Zhang, Yineng and Wang, Stephanie and Chen, Tianqi and Kasikci, Baris and Grover, Vinod and Krishnamurthy, Arvind and Ceze, Luis},
  booktitle = {Proceedings of Machine Learning and Systems (MLSys)},
  year      = {2025}
}

@inproceedings{wang2026nezha,
  title     = {{NEZHA}: A Zero-sacrifice and Hyperspeed Decoding Architecture for Generative Recommendations},
  author    = {Wang, Yejing and Zhou, Shengyu and Lu, Jinyu and Liu, Ziwei and Liu, Langming and Wang, Maolin and Zhang, Wenlin and Li, Feng and Su, Wenbo and Wang, Pengjie and Xu, Jian and Zhao, Xiangyu},
  booktitle = {Proceedings of the ACM Web Conference 2026},
  year      = {2026}
}

@article{shi2025lladarec,
  title   = {{LLaDA-Rec}: Discrete Diffusion for Parallel Semantic {ID} Generation in Generative Recommendation},
  author  = {Shi, Teng and Shen, Chenglei and Yu, Weijie and Nie, Shen and Li, Chongxuan and Zhang, Xiao and He, Ming and Han, Yan and Xu, Jun},
  journal = {arXiv preprint arXiv:2511.06254},
  year    = {2025}
}

@inproceedings{yu2023megabyte,
  title     = {{MEGABYTE}: Predicting Million-byte Sequences with Multiscale Transformers},
  author    = {Yu, Lili and Simig, D{\'a}niel and Flaherty, Colin and Aghajanyan, Armen and Zettlemoyer, Luke and Lewis, Mike},
  booktitle = {Advances in Neural Information Processing Systems},
  year      = {2023}
}

@inproceedings{ho2024block,
  title     = {Block Transformer: Global-to-Local Language Modeling for Fast Inference},
  author    = {Ho, Namgyu and Bae, Sangmin and Kim, Taehyeon and Jo, Hyunjik and Kim, Yireun and Schuster, Tal and Fisch, Adam and Thorne, James and Yun, Se-Young},
  booktitle = {Advances in Neural Information Processing Systems},
  year      = {2024}
}

@inproceedings{kasai2021deep,
  title     = {Deep Encoder, Shallow Decoder: Reevaluating Non-autoregressive Machine Translation},
  author    = {Kasai, Jungo and Pappas, Nikolaos and Peng, Hao and Cross, James and Smith, Noah A.},
  booktitle = {International Conference on Learning Representations},
  year      = {2021}
}

@article{barrault2024lcm,
  title   = {Large Concept Models: Language Modeling in a Sentence Representation Space},
  author  = {{LCM team} and Barrault, Lo{\"i}c and Duquenne, Paul-Ambroise and Elbayad, Maha and Kozhevnikov, Artyom and Alastruey, Belen and Andrews, Pierre and Coria, Mariano and Couairon, Guillaume and Costa-juss{\`a}, Marta R. and others},
  journal = {arXiv preprint arXiv:2412.08821},
  year    = {2024},
  doi     = {10.48550/arXiv.2412.08821}
}

@inproceedings{kim2016sequence,
  title     = {Sequence-Level Knowledge Distillation},
  author    = {Kim, Yoon and Rush, Alexander M.},
  booktitle = {Proceedings of the 2016 Conference on Empirical Methods in Natural Language Processing},
  pages     = {1317--1327},
  year      = {2016}
}

@inproceedings{tang2018ranking,
  title     = {Ranking Distillation: Learning Compact Ranking Models With High Performance for Recommender System},
  author    = {Tang, Jiaxi and Wang, Ke},
  booktitle = {Proceedings of the 24th ACM SIGKDD International Conference on Knowledge Discovery \& Data Mining},
  pages     = {2289--2298},
  year      = {2018}
}

@inproceedings{yuan2020revisiting,
  title     = {Revisiting Knowledge Distillation via Label Smoothing Regularization},
  author    = {Yuan, Li and Tay, Francis E. H. and Li, Guilin and Wang, Tao and Feng, Jiashi},
  booktitle = {Proceedings of the IEEE/CVF Conference on Computer Vision and Pattern Recognition},
  year      = {2020}
}

@inproceedings{vinyals2015pointer,
  title     = {Pointer Networks},
  author    = {Vinyals, Oriol and Fortunato, Meire and Jaitly, Navdeep},
  booktitle = {Advances in Neural Information Processing Systems},
  year      = {2015}
}

@inproceedings{see2017get,
  title     = {Get To The Point: Summarization with Pointer-Generator Networks},
  author    = {See, Abigail and Liu, Peter J. and Manning, Christopher D.},
  booktitle = {Proceedings of the 55th Annual Meeting of the Association for Computational Linguistics (Volume 1: Long Papers)},
  pages     = {1073--1083},
  year      = {2017}
}

@inproceedings{ren2019repeatnet,
  title     = {{RepeatNet}: A Repeat Aware Neural Recommendation Machine for Session-Based Recommendation},
  author    = {Ren, Pengjie and Chen, Zhumin and Li, Jing and Ren, Zhaochun and Ma, Jun and de Rijke, Maarten},
  booktitle = {Proceedings of the AAAI Conference on Artificial Intelligence},
  volume    = {33},
  pages     = {4806--4813},
  year      = {2019}
}

@article{qwen2025qwen25,
  title   = {Qwen2.5 Technical Report},
  author  = {{Qwen Team}},
  journal = {arXiv preprint arXiv:2412.15115},
  year    = {2024}
}

@inproceedings{hu2022lora,
  title     = {{LoRA}: Low-Rank Adaptation of Large Language Models},
  author    = {Hu, Edward J. and Shen, Yelong and Wallis, Phillip and Allen-Zhu, Zeyuan and Li, Yuanzhi and Wang, Shean and Wang, Lu and Chen, Weizhu},
  booktitle = {International Conference on Learning Representations},
  year      = {2022}
}

@article{wang2024multilingual,
  title   = {Multilingual {E5} Text Embeddings: A Technical Report},
  author  = {Wang, Liang and Yang, Nan and Huang, Xiaolong and Yang, Linjun and Majumder, Rangan and Wei, Furu},
  journal = {arXiv preprint arXiv:2402.05672},
  year    = {2024}
}

@article{cheng2026capsid,
  title   = {{CapsID}: Soft-Routed Variable-Length Semantic {IDs} for Generative Recommendation},
  author  = {Cheng, Wenzhuo and Gong, Menghang and Guo, Qixin and Zheng, Hang and Yang, Zhaobin and Lou, Jianguo and Zheng, Zhengwei},
  journal = {arXiv preprint arXiv:2605.05096},
  year    = {2026}
}

@inproceedings{he2025plum,
  title     = {{PLUM}: Adapting Pre-trained Language Models for Industrial-scale Generative Recommendations},
  author    = {He, Ruining and Heldt, Lukasz and Hong, Lichan and Keshavan, Raghunandan and Mao, Shifan and Mehta, Nikhil and Su, Zhengyang and Tsai, Alicia and Wang, Yueqi and Wang, Shao-Chuan and Yi, Xinyang and Baugher, Lexi and Cakici, Baykal and Chi, Ed and Goodrow, Cristos and Han, Ningren and Ma, He and Rosales, Romer and Van Soest, Abby and Tandon, Devansh and Wu, Su-Lin and Yang, Weilong and Zheng, Yilin},
  booktitle = {Proceedings of the ACM Web Conference 2026},
  pages     = {8093--8104},
  year      = {2026},
  doi       = {10.1145/3774904.3792802}
}

@article{kong2025minionerec,
  title   = {{MiniOneRec}: An Open-Source Framework for Scaling Generative Recommendation},
  author  = {Kong, Xiaoyu and Sheng, Leheng and Tan, Junfei and Chen, Yuxin and Wu, Jiancan and Zhang, An and Wang, Xiang and He, Xiangnan},
  journal = {arXiv preprint arXiv:2510.24431},
  year    = {2025}
}

@article{denadai2026neo,
  title   = {A Unified Language Model for Large Scale Search, Recommendation, and Reasoning},
  author  = {De Nadai, Marco and D'Amico, Edoardo and Lefarov, Max and Tamborrino, Alexandre and Vohra, Divita and VanMiddlesworth, Mark and Lin, Shawn and Wood, Jacqueline and Stypka, Jan and Klyce, Eliza and Dai, Keshi and Heath, Timothy Christopher and Gould, Martin D. and Raimond, Yves and Ghael, Sandeep and Jebara, Tony and Damianou, Andreas and Radosavljevic, Vladan and Bennett, Paul N. and Lalmas, Mounia and Chandar, Praveen},
  journal = {arXiv preprint arXiv:2603.17533},
  year    = {2026},
  doi     = {10.48550/arXiv.2603.17533}
}

@article{sun2025xgr,
  title   = {{xGR}: Efficient Generative Recommendation Serving at Scale},
  author  = {Sun, Qingxiao and Liu, Tongxuan and Zhang, Shen and Wu, Siyu and Yang, Peijun and Liang, Haotian and Li, Menxin and Ma, Xiaolong and Liang, Zhiwei and Ren, Ziyi and Zhang, Minchao and Wang, Yifan and Liu, Xinyu and Zhang, Ke and Yang, Hailong and Qian, Depei},
  journal = {arXiv preprint arXiv:2512.11529},
  year    = {2025}
}

@article{zhou2025onerecv2,
  title   = {{OneRec-V2} Technical Report},
  author  = {Zhou, Guorui and others},
  journal = {arXiv preprint arXiv:2508.20900},
  year    = {2025}
}

@article{dragunov2025sonar,
  title={{SONAR-LLM}: Autoregressive transformer that thinks in sentence embeddings and speaks in tokens},
  author={Dragunov, Nikita and Rahmatullaev, Temurbek and Goncharova, Elizaveta and Kurdiukov, Nikita and Mirzoeva, Aysel and Borisiuk, Anna and Kuznetsov, Andrey and Razzhigaev, Anton},
  journal={arXiv preprint arXiv:2508.05305},
  year={2025}
}

\appendix
\section{Additional Results}
\label{app:results}

\begin{figure*}[h]
  \setlength{\abovecaptionskip}{6pt}
  \centering
  \includegraphics[width=0.85\textwidth]{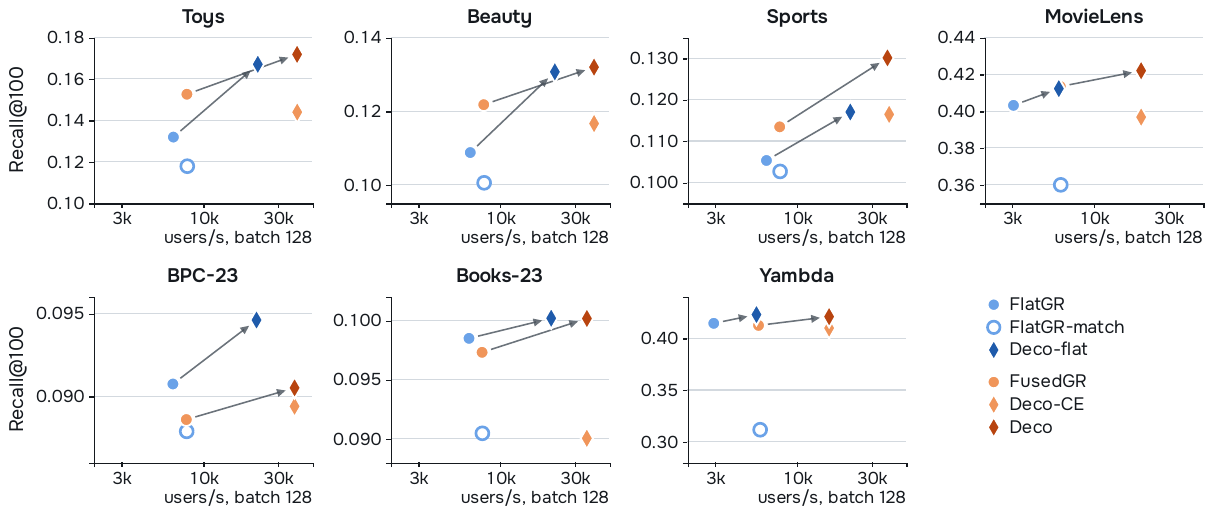}
  \caption{Accuracy--throughput trade-offs (Recall@100 against users per second at batch 128, log scale). Arrows connect full-backbone teachers to their concept-decoder students: \flatgr to \decoflat and \fusedgr to \deco. Both gain recall and throughput on every dataset.}
  \label{fig:acc-throughput-recall}
\end{figure*}

\paragraph{Accuracy against throughput per dataset.}
Figure~\ref{fig:acc-throughput-recall} places every model of Table~\ref{tab:main} in the accuracy--throughput plane.
The arrows connect each full-backbone teacher with its concept-decoder student.
On all seven datasets both arrows point to the right and upwards: concept decoding moves a model to higher throughput and higher Recall@100 at the same time.
The two other variants fall short in different directions.
\flatgrmatch is fast but loses accuracy, and \decoce is as fast as \deco but less accurate.

\begin{figure}[h]
  \setlength{\abovecaptionskip}{6pt}
  \centering
  \includegraphics[width=0.7\columnwidth]{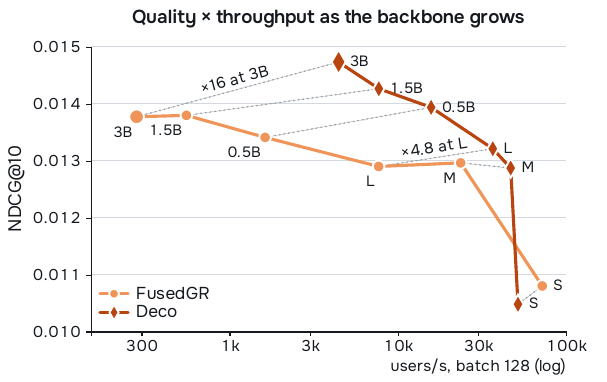}
  \caption{Quality against throughput as the backbone grows on BPC-23, from S, M and L (4M, 20M and 69M parameters, trained from scratch) to Qwen2.5 0.5B, 1.5B and 3B (fine-tuned with LoRA, merged for timing). \fusedgr slides left as the backbone grows (from 73k to 281 users/s), while \deco climbs in NDCG@10 and slows down far less (from 52k to 4.5k users/s).}
  \label{fig:backbone-tradeoff}
\end{figure}

\paragraph{Quality--throughput trajectories.}
Figure~\ref{fig:backbone-tradeoff} combines both panels of Figure~\ref{fig:backbone-scaling}.
As the backbone grows, \fusedgr moves almost horizontally to the left, buying small accuracy gains with large losses in throughput.
\deco climbs higher and slows down far less (12$\times$ against 258$\times$ from S to 3B), so every larger backbone gives a more accurate model at moderate cost.

\begin{figure}[h]
  \setlength{\abovecaptionskip}{5pt}
  \centering
  \includegraphics[width=0.90\columnwidth]{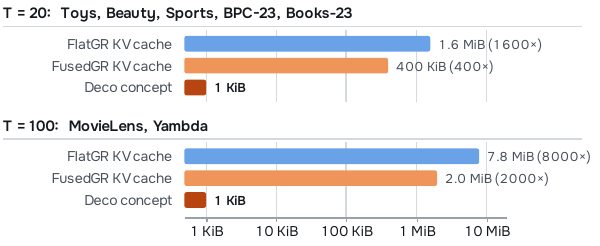}
  \caption{Per-user state to skip history encoding at request time (bf16, log scale): the backbone KV cache of the history versus the cached concept $u$. Beam-local decoder caches and the optional copy state are excluded.}
  \label{fig:per-user-state}
\end{figure}

\paragraph{Per-user serving state.}
Figure~\ref{fig:per-user-state} compares the state that must be kept per user to skip history encoding at request time: the KV cache of a full-backbone model grows with the history and its tokenization, whereas the \deco concept stays at 1~KiB.

\paragraph{Candidate budget.}
Table~\ref{tab:budget} complements Figure~\ref{fig:candidate-budget}.
A beam wider than $k$ adds little: about 3--4\% to Recall@100 and 2\% to Recall@300, and widening it from 300 to 1000 barely changes Recall@100.

\begin{table}[b]
  \setlength{\abovecaptionskip}{6pt}
  \caption{Candidate budget on BPC-23: relative gain in Recall@$k$ from a beam wider than $k$ (ref.: $B=k$).}
  \label{tab:budget}
  \footnotesize
  \begin{tabular*}{\columnwidth}{@{\extracolsep{\fill}} l ccc cc}
    \toprule
    & \multicolumn{3}{c}{\textbf{Recall@100}} & \multicolumn{2}{c}{\textbf{Recall@300}} \\
    \cmidrule(lr){2-4} \cmidrule(lr){5-6}
    \textbf{Beam width} & $B=100$ & $B=300$ & $B=1000$ & $B=300$ & $B=1000$ \\
    \midrule
    \flatgr & ref. & +2.9\% & +2.9\% & ref. & +2.0\% \\
    \fusedgr & ref. & +3.6\% & +3.6\% & ref. & +1.7\% \\
    \deco & ref. & +3.7\% & +3.8\% & ref. & +1.9\% \\
    \bottomrule
  \end{tabular*}
\end{table}

\paragraph{Decoder size.}
Table~\ref{tab:decoder-size} details the decoder-size study of \S\ref{sec:rq5}.
Across a sixfold range of decoder parameters, NDCG@10 and Recall@100 stay within about 3\% of the default, while latency grows with depth and width.

\begin{table}[t]
  \setlength{\abovecaptionskip}{6pt}
  \caption{Decoder size: \deco with light decoders of different depth and width (default shaded). Latency at batch 128, relative to the default.}
  \label{tab:decoder-size}
  \setlength{\tabcolsep}{3pt}
  \resizebox{\columnwidth}{!}{%
  \begin{tabular}{c r cc c cc c}
    \toprule
    & & \multicolumn{2}{c}{\textbf{Sports}} & & \multicolumn{2}{c}{\textbf{BPC-23}} & \\
    \cmidrule(lr){3-4} \cmidrule(lr){6-7}
    \textbf{Layers\,$\times$\,width} & \textbf{Params} & N@10 & R@100 & & N@10 & R@100 & \textbf{Latency}\,{\scriptsize$\downarrow$} \\
    \midrule
    1\,$\times$\,128 & 3.5M & 0.0192\pmsd{.0008} & 0.1297\pmsd{.0011} & & 0.0131\pmsd{.0002} & 0.0899\pmsd{.0007} & \xs0.72 \\
    2\,$\times$\,128 & 3.8M & 0.0194\pmsd{.0011} & 0.1308\pmsd{.0021} & & 0.0132\pmsd{.0000} & 0.0901\pmsd{.0005} & \xs0.84 \\
    1\,$\times$\,256 & 7.6M & 0.0186\pmsd{.0008} & 0.1276\pmsd{.0018} & & 0.0133\pmsd{.0001} & 0.0909\pmsd{.0003} & \xs0.80 \\
    \rowcolor{decobg} 2\,$\times$\,256 & 8.6M & 0.0191\pmsd{.0007} & 0.1301\pmsd{.0023} & & 0.0132\pmsd{.0003} & 0.0905\pmsd{.0005} & \multicolumn{1}{c}{\footnotesize ref.} \\
    3\,$\times$\,256 & 9.7M & 0.0194\pmsd{.0012} & 0.1306\pmsd{.0017} & & 0.0133\pmsd{.0002} & 0.0905\pmsd{.0006} & \xs1.20 \\
    4\,$\times$\,256 & 10.7M & 0.0191\pmsd{.0010} & 0.1300\pmsd{.0001} & & 0.0133\pmsd{.0002} & 0.0899\pmsd{.0008} & \xs1.39 \\
    2\,$\times$\,512 & 21.6M & 0.0186\pmsd{.0016} & 0.1260\pmsd{.0020} & & 0.0131\pmsd{.0002} & 0.0899\pmsd{.0004} & \xs1.46 \\
    \bottomrule
  \end{tabular}}
\end{table}

\paragraph{Distillation without the concept decoder.}
To separate the concept decoder from distillation, we also distill the \fusedgr teacher into a \fusedgr student with the objective and loss weighting of \deco (\fusedgr + KD, Table~\ref{tab:selfkd}).
Its NDCG@10 gain over the teacher is less than half that of \deco: 7.5\% vs.~19.4\% on Sports and 0.8\% vs.~2.3\% on BPC-23.
Since the concept decoder alone (\decoce) stays close to \fusedgr, the two reinforce each other: the teacher's graded preferences are worth more to the light decoder than to a full-backbone student.

\begin{table}[t]
  \setlength{\abovecaptionskip}{6pt}
  \caption{Distillation with and without the concept decoder. \fusedgr + KD: a \fusedgr student distilled from the same teacher as \deco.}
  \label{tab:selfkd}
  \setlength{\tabcolsep}{7pt}
  \resizebox{\columnwidth}{!}{%
  \begin{tabular}{l cc c cc}
    \toprule
    & \multicolumn{2}{c}{\textbf{Sports}} & & \multicolumn{2}{c}{\textbf{BPC-23}} \\
    \cmidrule(lr){2-3} \cmidrule(lr){5-6}
    \textbf{Model} & N@10 & R@100 & & N@10 & R@100 \\
    \midrule
    \fusedgr (teacher) & 0.0160\pmsd{.0003} & 0.1134\pmsd{.0008} & & 0.0129\pmsd{.0001} & 0.0886\pmsd{.0003} \\
    \fusedgr + KD & 0.0172\pmsd{.0014} & 0.1196\pmsd{.0005} & & 0.0130\pmsd{.0004} & 0.0896\pmsd{.0005} \\
    \addlinespace[2pt]
    \decoce & 0.0164\pmsd{.0004} & 0.1164\pmsd{.0035} & & 0.0127\pmsd{.0001} & 0.0894\pmsd{.0007} \\
    \rowcolor{decobg} \deco & 0.0191\pmsd{.0007} & 0.1301\pmsd{.0023} & & 0.0132\pmsd{.0003} & 0.0905\pmsd{.0005} \\
    \bottomrule
  \end{tabular}}
\end{table}

\section{Tied and Untied Code Tables}
\label{app:tying}

Code vectors play up to three roles: they describe history items (through the fuser or as flat input tokens), they condition the decoder on the generated prefix, and they score the next code as output class vectors.
All main experiments keep separate tables for these roles, the more general choice: a code can have different learned representations for reading and writing, and extra item features can enter the input side without changing the output classes.
In Table~\ref{tab:tying}, we tie them into one table per level, from which the history fuser reads RMS-normalized rows.
Tying raises mean Recall@100 for \flatgr, \fusedgr and \deco on both datasets, while the effect on \decoce is mixed.
The \emph{Content} column shows what tying changes.
First-level codes are clusters of similar items, yet in the untied fused models the ten nearest neighbors of a code among the history-side vectors barely overlap with its ten nearest clusters, and these vectors keep the norm of their random initialization.
The output class vectors, in contrast, recover the content neighborhoods well above chance (8--11\% on Sports, 39--41\% on Yambda), and tying shares them with the history side; \flatgr inputs partially recover these neighborhoods even with untied tables.
The main comparison does not depend on this choice: \deco is the most accurate of the four models with both tied and untied tables.

\begin{table}[t]
  \setlength{\abovecaptionskip}{6pt}
  \caption{Untied vs.\ tied SID code tables: Recall@100 and \mbox{Content} (untied\,$\to$\,tied), the kNN@10 overlap (\%) of a first-level code's cosine neighborhoods among the history-side code vectors and among the RQ-KMeans centroids (codes seen at least 20 times in training; chance 0.5--0.6\%).}
  \label{tab:tying}
  \setlength{\tabcolsep}{3pt}
  \resizebox{\columnwidth}{!}{%
  \begin{tabular}{l ccr@{\hspace{15pt}}ccr}
    \toprule
    & \multicolumn{3}{c}{\textbf{Sports}} & \multicolumn{3}{c}{\textbf{Yambda}} \\
    \cmidrule(lr){2-4} \cmidrule(lr){5-7}
    \textbf{Model} & Untied & Tied & Content, \% & Untied & Tied & Content, \% \\
    \midrule
    \flatgr & 0.1053\pmsd{.0036} & 0.1128\pmsd{.0030} & 6.1\,$\to$\,10.4 & 0.4144\pmsd{.0022} & 0.4162\pmsd{.0012} & 25.9\,$\to$\,41.3 \\
    \fusedgr & 0.1134\pmsd{.0008} & 0.1212\pmsd{.0021} & 0.7\,$\to$\,10.6 & 0.4125\pmsd{.0025} & 0.4183\pmsd{.0015} & 1.5\,$\to$\,46.1 \\
    \decoce & 0.1164\pmsd{.0035} & 0.1129\pmsd{.0044} & 0.7\,$\to$\,\phantom{0}9.9 & 0.4099\pmsd{.0016} & 0.4128\pmsd{.0012} & 1.5\,$\to$\,43.7 \\
    \deco & 0.1301\pmsd{.0023} & 0.1317\pmsd{.0004} & 0.7\,$\to$\,12.6 & 0.4209\pmsd{.0007} & 0.4254\pmsd{.0005} & 1.6\,$\to$\,45.7 \\
    \bottomrule
  \end{tabular}}
\end{table}

\section{Further Implementation Details}
\label{app:impl}

\paragraph{Architectures.}
Table~\ref{tab:models} summarizes how each model reads the history and where its beam search runs.
The scratch backbones S, M and L have widths 128, 384 and 640, feed-forward sizes 512, 1536 and 2560, and 2, 4 and 8 layers with 2, 6 and 10 heads.
L is the default backbone of all other experiments.
The Qwen2.5 backbones read the same fused item tokens as input embeddings.
The fuser embeds each code in 64 dimensions (160 on Yambda and Books-23).
The light decoder (2 layers, width 256, 4 heads, $M=2$) is the same for every backbone; the decoder-size study of \S\ref{sec:rq5} varies its depth from 1 to 4 layers and its width from 128 to 512 (Table~\ref{tab:decoder-size}).

\begin{table}[t]
  \setlength{\abovecaptionskip}{6pt}
  \caption{How the backbone reads the history and where beam search runs in each model. The light decoder sees only the concept $z$ and the generated prefix. $T$ is the number of history items (the last $T/K$ for \flatgrmatch); $K$ is the number of codes per SID.}
  \label{tab:models}
  \centering
  \footnotesize
  \setlength{\tabcolsep}{8.8pt}
  \begin{tabular}{l c c l l}
    \toprule
    \textbf{Model} & \begin{tabular}[b]{@{}c@{}}\textbf{Tokens}\\\textbf{per item}\end{tabular} & \begin{tabular}[b]{@{}c@{}}\textbf{Input}\\\textbf{length}\end{tabular} & \textbf{Decoder} & \textbf{Training} \\
    \midrule
    \flatgr      & $K$ & $TK$ & backbone & CE \\
    \flatgrmatch & $K$ & $T$  & backbone & CE \\
    \fusedgr   & $1$ & $T$  & backbone & CE \\
    \midrule
    \decoflat  & $K$ & $TK$ & light decoder & CE+KD \\
    \decoce    & $1$ & $T$  & light decoder & CE \\
    \rowcolor{decobg}\deco & $1$ & $T$ & light decoder & CE+KD \\
    \bottomrule
  \end{tabular}
\end{table}

\paragraph{Training.}
Models are trained for at most 5k optimizer steps on the Amazon-2014 datasets, 20k on MovieLens, 30k on BPC-23 and Books-23 and 60k on Yambda, with early stopping on validation NDCG@10.
Scratch-model learning rates are selected per family and dataset on validation, in the range $10^{-4}$--$10^{-3}$, with backbone dropout 0.3.
The distillation weights in Eq.~\eqref{eq:loss} are set once from the initial values $\bar\ell_{\mathrm{ce}}$ and $\bar\ell_{\mathrm{kd}}$ of the two terms on a fixed sample, $w_{\mathrm{ce}}=1/(3\bar\ell_{\mathrm{ce}})$ and $w_{\mathrm{kd}}=2/(3\bar\ell_{\mathrm{kd}})$, so only the temperature is set per dataset: $\tau=1$, and $\tau=2$ on Beauty and MovieLens.
The Qwen2.5 backbones use family-specific recipes selected on validation.

\paragraph{Timing.}
All efficiency results use one H100 80GB GPU and beam width 100 unless stated otherwise.
Our six models and the scaling studies use one serving pipeline for every model family.
Weights are in bf16.
Each user's history keys and values form one beam-shared history cache; attention over this prefix runs once for all beams and is merged with each beam's own suffix, as in shared-prefix attention for LLM serving~\cite{juravsky2024hydragen,ye2025flashinfer}.
Normalization, feed-forward and beam-step operations are compiled with \texttt{torch.compile}, and the prefill and the beam search are captured as two CUDA graphs per batch shape.
Compilation, graph capture and warm-up are not timed.
Each point runs 10 warm-up and 100 timed batches of test histories (5 and 50 above batch 64) and is repeated three times in a shuffled order.
Latency covers the backbone prefill and the full beam search for a batch, including trie masking; data loading and the mapping from SIDs to items are excluded.
Throughput is the batch size divided by the median batch latency.
The batch-size sweep doubles the batch from 1 until throughput levels off, and the peak is reported at the smallest batch within 3\% of the highest throughput.
For the p95 budget in Figure~\ref{fig:serving}, we report the highest throughput among the batches whose p95 latency stays within 50\,ms; batch 128 is the largest batch at which every full-backbone baseline meets it on every dataset, at 88--97\% of its peak throughput.
Under this budget, \deco serves 3.2--7.3$\times$ more users than \fusedgr on all seven datasets (6.5--8.6$\times$ more than \flatgr).
Table~\ref{tab:baselines} times every model through its own retrieval path in eager PyTorch (bf16 autocast, batch 128).

\end{document}